\documentclass[fleqn,usenatbib]{mnras}

\usepackage{newtxtext}
\usepackage[T1]{fontenc}

\DeclareRobustCommand{\VAN}[3]{#2}
\let\VANthebibliography\thebibliography
\def\thebibliography{\DeclareRobustCommand{\VAN}[3]{##3}\VANthebibliography}

\usepackage{amsmath}	% Advanced maths commands
\usepackage{amssymb}	% Extra maths symbols

\usepackage{graphicx}
\usepackage{txfonts}
\usepackage{hyperref}
\usepackage{xcolor}
\usepackage[shortlabels]{enumitem}
\usepackage{tablefootnote}
\usepackage{ulem}

\title{Mapper of the IGM spin temperature: instrument overview}

\author[R. A. Monsalve et al.]{
R. A. Monsalve,$^{1,2,3}$\thanks{E-mail: raul.monsalve@berkeley.edu}\thanks{Sadly, Mauricio D\'iaz passed away before the publication of this article.}
C. Altamirano,$^{3}$
V. Bidula,$^{4}$
R. Bustos,$^{3}$
C. H. Bye,$^{5}$
H. C. Chiang,$^{4,6}$\newauthor
M. D\'iaz,$^{3}$
B. Fern\'andez,$^{3}$
X. Guo,$^{1}$
I. Hendricksen,$^{4}$
E. Hornecker,$^{7}$
F. Lucero,$^{8}$
H. Mani,$^{9}$\newauthor
F. McGee,$^{4}$
F. P. Mena,$^{10}$
M. Pess\^oa,$^{4}$
G. Prabhakar,$^{1}$
O. Restrepo,$^{8,11}$
J. L. Sievers,$^{4,12}$
\and N. Thyagarajan,$^{13}$\newauthor
\\
$^{1}$Space Sciences Laboratory, University of California, Berkeley, CA 94720, USA\\
$^{2}$School of Earth and Space Exploration, Arizona State University, Tempe, AZ 85287, USA\\
$^{3}$Departamento de Ingenier\'ia El\'ectrica, Universidad Cat\'olica de la Sant\'isima Concepci\'on, Alonso de Ribera 2850, Concepci\'on, Chile\\
$^{4}$Department of Physics and Trottier Space Institute, McGill University, Montr\'eal, QC H3A 2T8, Canada\\
$^{5}$Department of Astronomy, University of California, Berkeley, CA 94720, USA\\
$^{6}$School of Mathematics, Statistics, \& Computer Science, University of KwaZulu-Natal, Durban, South Africa\\
$^{7}$David A. Dunlap Department of Astronomy \& Astrophysics, University of Toronto, Toronto, ON M5S 3H4, Canada\\
$^{8}$Departamento de Ingenier\'ia El\'ectrica, Universidad de Chile, Santiago, Chile\\
$^{9}$CryoElec LLC, Chandler, AZ 85225, USA\\
$^{10}$National Radio Astronomy Observatory, Charlottesville, VA 22903, USA\\
$^{11}$Facultad de Ingenier\'ia, Universidad ECCI, Bogot\'a, 111311, Colombia\\
$^{12}$School of Chemistry and Physics, University of KwaZulu-Natal, Durban, South Africa\\
$^{13}$Commonwealth Scientific and Industrial Research Organisation (CSIRO), Space \& Astronomy, P. O. Box 1130, Bentley, WA 6102, Australia\\
}

\date{Received XXXX; accepted XXXX}
\pubyear{2023}

\begin{document}
\label{firstpage}
\pagerange{\pageref{firstpage}--\pageref{lastpage}}
\maketitle

\begin{abstract}
The observation of the global $21$~cm signal produced by neutral hydrogen gas in the intergalactic medium (IGM) during the Dark Ages, Cosmic Dawn, and Epoch of Reionization requires measurements with extremely well-calibrated wideband radiometers. We describe the design and characterization of the Mapper of the IGM Spin Temperature (MIST), which is a new ground-based, single-antenna, global $21$~cm experiment. The design of MIST was guided by the objectives of avoiding systematics from an antenna ground plane and cables around the antenna, as well as maximizing the instrument's on-sky efficiency and portability for operations at remote sites. We have built two MIST instruments, which observe in the range $25$--$105$~MHz. For the $21$~cm signal, this frequency range approximately corresponds to redshifts $55.5 > z > 12.5$, encompassing the Dark Ages and Cosmic Dawn. The MIST antenna is a horizontal blade dipole of $2.42$~m in length, $60$~cm in width, and $52$~cm in height above the ground. This antenna operates without a metal ground plane. The instruments run on $12$~V batteries and have a maximum power consumption of $17$~W. The batteries and electronics are contained in a single receiver box located under the antenna. We present the characterization of the instruments using electromagnetic simulations and lab measurements. We also show sample sky measurements from recent observations at remote sites in California, Nevada, and the Canadian High Arctic. These measurements indicate that the instruments perform as expected. Detailed analyses of the sky measurements are left for future work.
\end{abstract}

\begin{keywords}
astronomical instrumentation, methods and techniques -- instrumentation: miscellaneous -- methods: observational -- dark ages, reionization, first stars -- cosmology: observations. 
\end{keywords}

%
%-------------------------------------------------------------------
\section{Introduction}

The measurement of the $21$~cm line from neutral hydrogen gas in the intergalactic medium (IGM) has been recognized as a promising way to map the evolution of the Universe during its first billion years and as the only way to observe the Dark Ages \citep{hogan1979,madau1997,shaver1999,rees2000,tozzi2000,barkana2001,loeb2004,furlanetto2006a}.

The observation frequency of the $21$~cm signal from redshift $z$ is given by $\nu=1420\;\mathrm{MHz} \times (1+z)^{-1}$. Therefore, measurements of neutral hydrogen before the end of the Epoch of Reionization (EoR) have to be conducted at $\nu<220$~MHz\footnote{The Epoch of Reionization is currently estimated to have ended by $z\approx5.5$ \citep{kulkarni2019,nasir2020,cain2021,qin2021,raste2021}.}. Several radio experiments are trying to detect this cosmological signal. They can be classified into those targetting the sky-averaged or global component, and antenna arrays focussing on spatial anisotropies. Ground-based global $21$~cm experiments include ASSASSIN \citep{mckinley2020}, CTP \citep{nhan2019}, EDGES \citep{monsalve2017b,bowman2018}, LEDA \citep{bernardi2016,price2018}, LWA-SV \citep{dilullo2020}, PRI$^Z$M \citep{philip2019}, REACH \citep{deleraacedo2022,razavi2023}, SARAS \citep{singh2017,singh2022}, and SITARA \citep{thekkeppattu2022}. Space-based global signal missions and concepts include DAPPER \citep{burns2021}, DSL/Hongmeng \citep{shi2022b}, LuSEE~Night \citep{bale2023}, and PRATUSH.\footnote{\url{https://wwws.rri.res.in/DISTORTION/pratush.html}} Arrays targeting the $21$~cm spatial anisotropies from the ground include HERA \citep{deboer2017}, LOFAR \citep{vanhaarlem2013}, MWA \citep{tingay2013}, NenuFAR \citep{zarka2020}, OVRO-LWA \citep{garsden2021}, and SKA-Low \citep{koopmans2015}. Space-based array concepts include CoDEX \citep{koopmans2021}, the DSL/Hongmeng array \citep{chen2021,shi2022a}, FARSIDE \citep{burns2021}, and OLFAR \citep{bentum2020}.

In this paper, we introduce the Mapper of the IGM Spin Temperature (MIST),\footnote{\url{www.physics.mcgill.ca/mist}} which is a new ground-based experiment trying to detect the global $21$~cm signal. The brightness temperature of this signal is given by \citep{furlanetto2006a}

\begin{equation}
T_b(z) \approx 9 x_H(z) \left[1-\frac{T_{cmb}(z)}{T_{spin}(z)}\right]\sqrt{1+z}\;\;\;\mathrm{mK},
\end{equation}

\noindent where $x_H$ is the mean hydrogen neutral fraction, $T_{cmb}$ is the temperature of the cosmic microwave background, and $T_{spin}$ is the $21$~cm spin temperature. $T_b$ can be represented as a frequency spectrum using the redshift-to-frequency mapping for the $21$~cm line. Two absorption features are expected in this spectrum: one from the Dark Ages and the other from the Cosmic Dawn. The Dark Ages feature has a centre at $\approx17$~MHz ($z\approx80$), an amplitude of $\approx40$~mK, and a full width at half maximum (FWHM) of $\approx25$~MHz \citep{furlanetto2006a,pritchard2008,mondal2023}. The absorption feature from the Cosmic Dawn is centred somewhere in the range $\approx45$--$130$~MHz ($30\gtrsim z \gtrsim 10$) and has an amplitude of up to $\approx250$~mK. The exact shape of the Cosmic Dawn feature depends on the astrophysical characteristics of the first stars, galaxies, and black holes \citep{tozzi2000,furlanetto2006b,pritchard2010,cohen2017,mirocha2018}.

One of the main challenges in the global $21$~cm measurement is the presence of diffuse foreground contamination. This contamination is at least four orders of magnitude stronger than the signal, and dominated by Galactic and extragalactic synchrotron radiation \citep[e.g.,][]{voytek2014, bernardi2016}. Radio point sources that also affect this measurement include Cas~A, Cyg~A, Tau~A, and Vir~A \citep{helmboldt2009,degasperin2020}, while Jupiter and the Sun can have a significant transient contribution below $\approx40$~MHz \citep[e.g.,][]{panchenko2013,sasikumarraja2022}. Other serious challenges to this measurement include very stringent instrument calibration requirements \citep{monsalve2017a}, human-made radio-frequency interference (RFI) \citep{offringa2015,dyson2021}, and absorption, emission, and refraction from the Earth's ionosphere \citep{vedantham2014,rogers2015,sokolowski2015b,datta2016,shen2021}.

In \citet{bowman2018}, the EDGES global $21$~cm experiment made the only detection claim of the Cosmic Dawn absorption feature to date. The reported signal has a flattened Gaussian shape, an amplitude of $0.5_{-0.2}^{+0.5}$~K, a centre at $78\pm1$~MHz, and a FWHM of $19_{-2}^{+4}$~MHz. The ranges in each parameter represent $\pm3\sigma$ uncertainty, which is mainly systematic. The reported best-fit amplitude is at least twice as large as expected in standard models, which has motivated theorists to propose physical scenarios for the early Universe not typically considered before the EDGES result \citep[e.g.,][]{munoz2018,fialkov2019,mirocha2019}. However, the large absorption and unexpected flattened Gaussian shape have also produced skepticism about the cosmological interpretation of the feature \citep{hills2018,singh2019,sims2020}. Alternative explanations that have been suggested include instrumental systematics, such as resonances in the metal ground plane under the EDGES antenna \citep{bradley2019}, and contributions from other sources in the sky, such as polarized Galactic emission \citep{spinelli2019}. The SARAS3 experiment recently reported a non-detection of the EDGES signal using measurements with a vertical monopole antenna floating in a lake \citep{singh2022}. This null result has a significance of $<2\sigma$ when considering both the uncertainty of SARAS3 as well as the uncertainty of the reported EDGES signal, but it nonetheless increases the pressure to determine the origin of the EDGES feature. Whether or not the EDGES signal is cosmological, detecting and validating the global $21$~cm signal from the Dark Ages and Cosmic Dawn will require measurements from different experiments.

The MIST instrument is a single-antenna, single-polarization, total-power radiometer that observes the sky at $25$--$105$~MHz, which for the $21$~cm signal corresponds to redshifts $55.5 > z > 12.5$. This range represents a large fraction of the range where the Dark Ages and Cosmic Dawn absorption features are expected to be found. The instrument has been designed to achieve the performance required for detection while remaining highly portable for transportation to remote radio-quiet sites. A significant difference between MIST and other wideband-dipole, total-power radiometers, such as EDGES, LEDA, PRI$^Z$M, and REACH, is its operation without a metal ground plane over the soil. This choice eliminates systematic effects associated with finite ground planes and their physical limitations, with the tradeoff of folding soil properties into MIST's electromagnetic performance. The unique instrumental approach, combined with observations from multiple locations and terrestrial latitudes, will enable MIST to significantly contribute to the detection of the global $21$~cm signal through independent measurements subject to different experimental factors. 

We have built two copies of the MIST instrument, which we deployed in the field for the first time in 2022. Specifically, we conducted observations from Deep Springs Valley in California, the Sarcobatus Flat in Nevada, and the McGill Arctic Research Station (MARS) in the Canadian High Arctic.

This paper is organized as follows. Section~\ref{section_description} provides a general description of the MIST instruments. Section~\ref{section_formalism} describes the calibration formalism for the sky measurements. In Section~\ref{section_antenna} we show the characteristics of the MIST antenna and discuss its sensitivity to the electrical properties of the soil. In Section~\ref{section_balun} we provide details about the balun, which acts as an interface between the antenna and the receiver. Section~\ref{section_receiver} describes the receiver electronics and laboratory calibration. In Section~\ref{section_field_measurements} we present sample field measurements that provide an initial view of the instrument's performance. We summarise this paper in Section~\ref{section_summary}.

\begin{figure*}
\centering
\includegraphics[width=\linewidth]{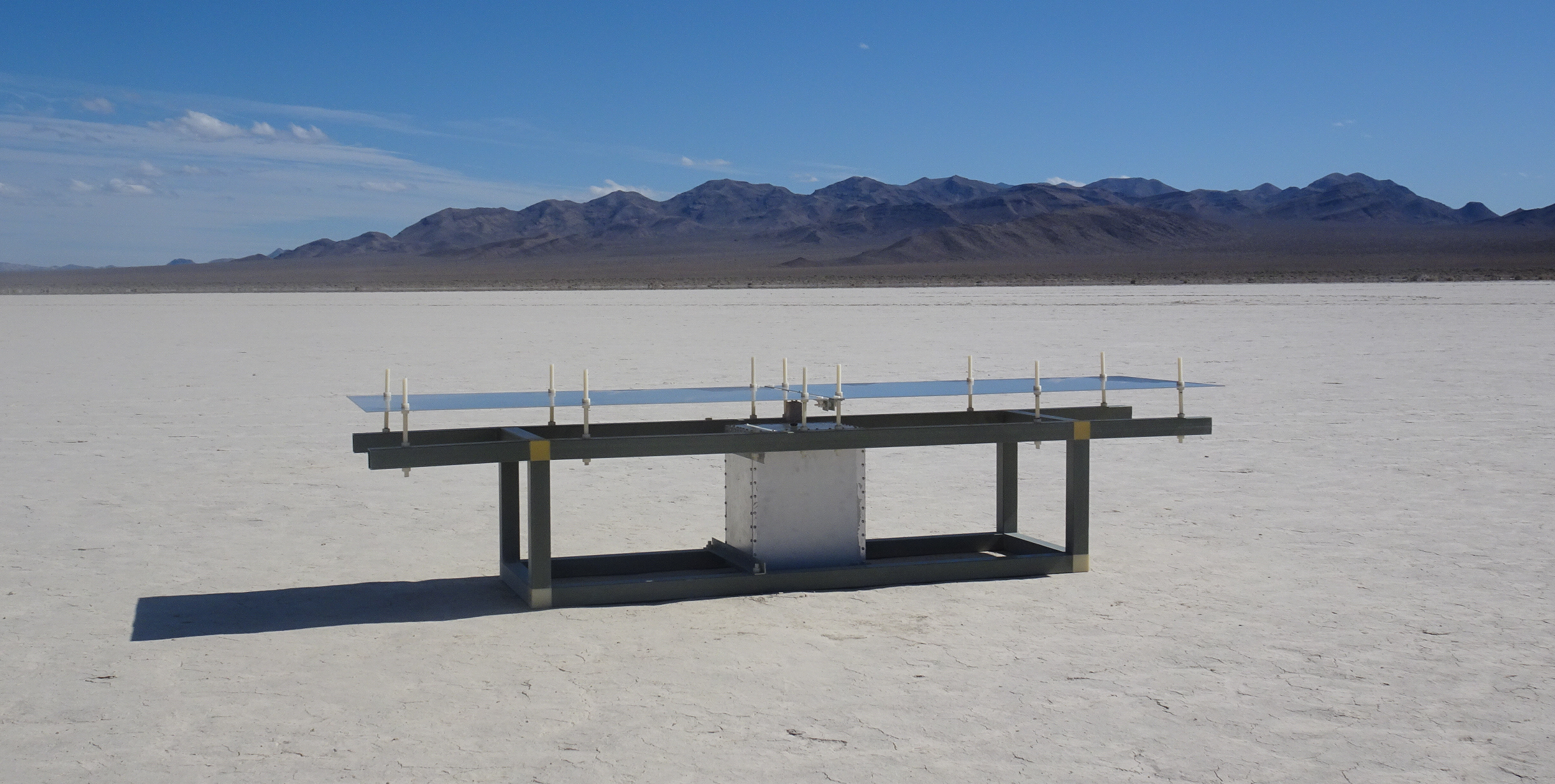}
\caption{One of the two MIST instruments during sky observations in the Sarcobatus Flat, Nevada, in May 2022.}
   \label{figure_death_valley}
\end{figure*}

%--------------------------------
\section{General description}
\label{section_description}
%--------------------------------
The MIST instrument\footnote{In this paper we refer to the two MIST instruments as `instrument' because their design is nominally identical.} is a single-antenna, single-polarization, total-power radiometer. The design of the instrument is motivated by the following objectives:

\begin{enumerate}[(1)]
\item Have an antenna beam pattern above the soil that peaks at the zenith and decreases toward the horizon. This pattern would maximize the sensitivity of the instrument to the sky signal, and minimize the sensitivity to features of the terrain and sky blockage by mountains \citep{bassett2021, pattison2023}.

\item Avoid using a metal ground plane. In addition to reducing the probability of instrumental resonances \citep{bradley2019}, this choice would eliminate the possibility of signal reflections produced by the electrical discontinuity between the edges of the ground plane and the soil \citep{mahesh2021,rogers2022,spinelli2022}. Operating without a ground plane increases the ground loss. However, the ground loss can in principle be estimated using electromagnetic simulations and then removed during data analysis. Observations from different sites could be leveraged to test for systematic effects related to ground loss.

\item Keep the instrument small and of low power consumption. These characteristics would enable the operation of the instrument with small batteries and facilitate its transportation to remote sites. In a small instrument, the radio-frequency (RF) paths would be short, increasing the accuracy with which key calibration parameters, in particular of the receiver, could be measured.

\item Avoid using cables outside of the receiver box. By avoiding cables we would eliminate their potential impact on the antenna performance, reduce emission of self-generated RFI, and simplify the design for electromagnetic simulations.
\end{enumerate}

From the guidelines above, we arrived at a design based on a horizontal $2.42$~m tip-to-tip blade dipole antenna made of solid aluminum panels. The antenna operates $52$~cm above the soil and without a metal ground plane. The instrument is powered by four $12$~V, $18$~Ah batteries, and has a maximum power consumption of $17$~W. Except for a small balun, all the electronics of the instrument, including the batteries, are contained in a single aluminum receiver box of size $40.5$~cm~$\times$~$33.5$~cm~$\times$~$26$~cm located under the antenna. The antenna and receiver box are supported by a frame made of fiberglass tubes and angles, acrylonitrile butadiene styrene (ABS) plastic elbows, and nylon rods, screws, and nuts. The full frequency range of the radiometer is $0$--$125$~MHz but the sky observations are limited to the range $25$--$105$~MHz. The high-frequency end of this range is imposed by the reduced performance of the analog-to-digital converter (ADC) above $105$~MHz. The low-frequency end of the range is imposed by the reduced efficiency of the antenna combined with a significant increase in the shortwave RFI below $25$~MHz at most locations. The average FWHM of the antenna beam directivity across $25$--$105$~MHz is $\approx85^{\circ}$.

\begin{figure}
\centering
\includegraphics[width=\linewidth]{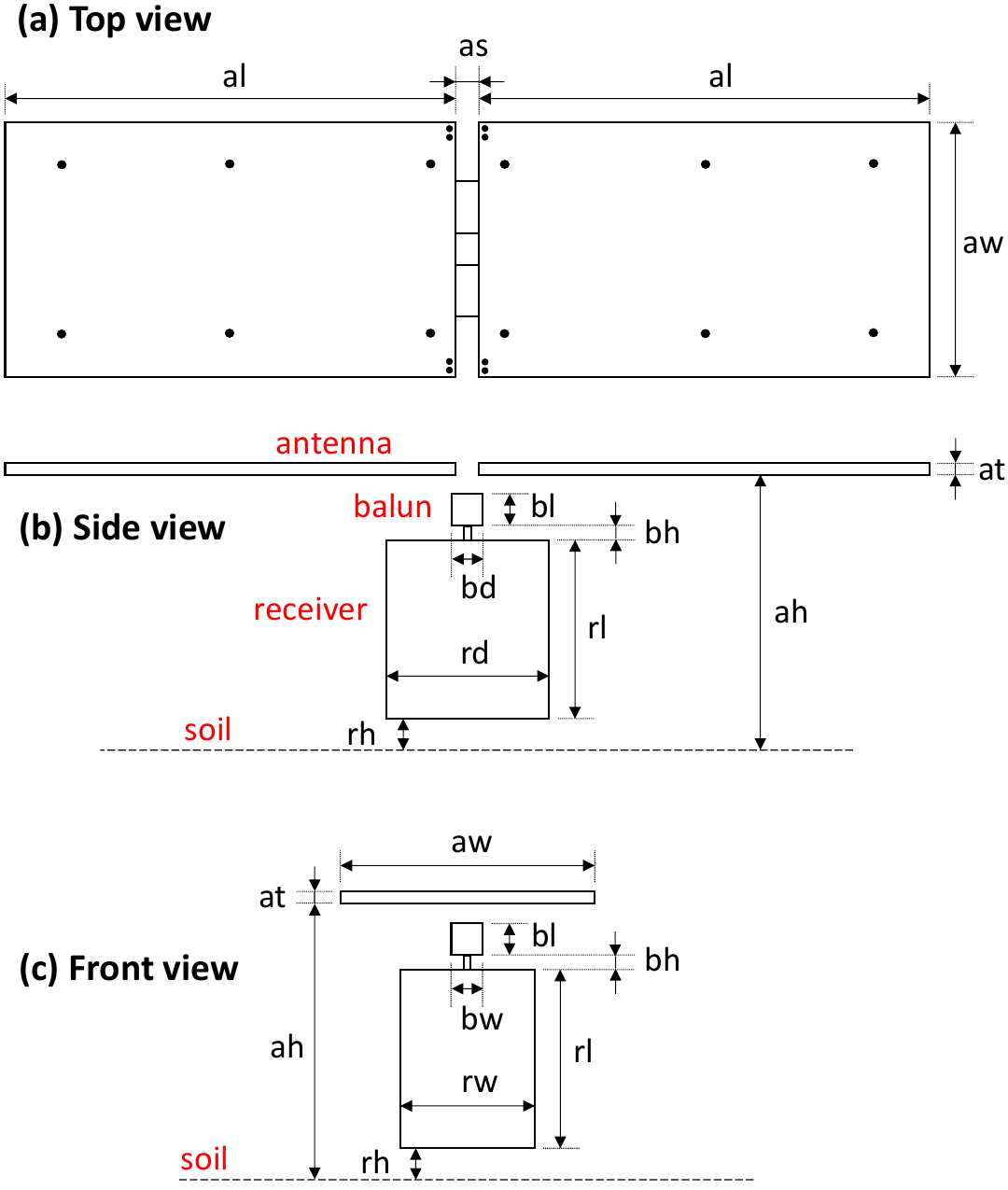}
\caption{Schematic of the MIST instrument (not to scale). The dimensions are listed in Table~\ref{table_antenna_dimensions}.}
\label{figure_instrument_diagram}
\end{figure}

Figure~\ref{figure_death_valley} shows a picture of one of the instruments deployed in the Sarcobatus Flat in May 2022. Figure~\ref{figure_instrument_diagram} shows a schematic of the instrument and Table~\ref{table_antenna_dimensions} lists the instrument's dimensions.

\section{Calibration formalism}
\label{section_formalism}

Before describing the instrument in detail, here we summarize the mathematical model used by MIST for the sky measurements.

\subsection{Sky contribution to the antenna temperature}
\label{section_formalism_antenna}

The contribution from the sky to the antenna temperature is given by

\begin{align}
T_S(\nu) = \frac{\int_0^{2\pi} \int_{0}^{\pi/2} T_{sky}(\theta,\phi,\nu)D(\theta,\phi,\nu) \sin\theta d \theta d\phi}{\int_0^{2\pi} \int_{0}^{\pi/2} D(\theta,\phi,\nu) \sin\theta d \theta d\phi},\label{equation_antenna_temperature2}
\end{align}

\noindent where $T_{sky}$ is the sky brightness temperature spatial distribution; $D$ is the antenna beam directivity; $\theta$ and $\phi$ are the antenna zenith and azimuth angles, respectively; and $\nu$ is frequency.

The antenna temperature measured at the input of the calibrated receiver is modeled as

\begin{align}
T_A(\nu) =  \eta(\nu)T_{S}(\nu) + \left[ 1-\eta(\nu) \right]T_{phys},\label{equation_antenna_temperature1}
\end{align}

\noindent where $\eta$ is the efficiency in the measurement of $T_S$ accounting for passive sources of loss, and $T_{phys}$ is the physical temperature associated with the passive sources of loss. MIST's on-sky efficiency corresponds to the multiplication of three factors: (1) the radiation efficiency, $\eta_{rad}$, which accounts for resistive loss due to finite electrical conductivity in and around the antenna; (2) the beam efficiency, $\eta_{beam}$, which corresponds to the fraction of the beam solid angle toward the sky relative to the total; and (3) the balun efficiency, $\eta_{balun}$, representing the efficiency of the signal transmission through the balun used between the antenna excitation port and the receiver input. Assuming the same physical temperature for the three passive sources of loss, which typically corresponds to the ambient temperature, Equation~\ref{equation_antenna_temperature1} can be solved for the sky contribution to obtain

\begin{align}
T_{S} = \frac{T_A -  \left[ 1- \eta_{rad}\eta_{beam}\eta_{balun} \right]T_{phys}}{\eta_{rad}\eta_{beam}\eta_{balun}}.
\end{align}

\subsection{Receiver calibration}
\label{section_formalism_receiver}

The antenna temperature calibrated at the receiver input is related to the power spectral density (PSD) measured by the receiver, $P_A$, by

\begin{equation}
T_A(\nu) = \frac{P_A(\nu)}{g_R(\nu)} - T_R(\nu),
\label{equation_temperature_psd}
\end{equation}

\noindent where $g_R$ is the receiver gain and $T_R$ is the receiver temperature. We determine $g_R$ and $T_R$ with the method developed in \citet{rogers2012} and \citet{monsalve2017a}. In this method, the receiver input is continuously switched in the field between three positions: (1) the antenna, (2) an internal ambient load, and (3) the internal ambient load in series with an active noise source. In addition, external calibration standards of different noise temperatures and reflection coefficients are measured in the lab. These external standards provide the absolute calibration to the receiver. With some rearrangement of the PSD equations for the three receiver input positions presented in \citet{monsalve2017a}, $g_R$ and $T_R$ can be written as (dropping the frequency dependences of all the quantities for brevity)

\begin{align}
g_R &= \left(\frac{1}{K_0}\right)\left[\frac{P_{L+NS} - P_L}{C_1 \left(T^a_{L+NS} - T^a_{L}\right)}\right],
\label{equation_full_receiver_gain}
\end{align}

\begin{align}
T_R = &K_0 \Biggl\{ P_L\left[\frac{C_1 \left(T^a_{L+NS} - T^a_{L}\right) }{P_{L+NS}-P_L}\right] - (T^a_L - C_2) \Biggl\}\nonumber\\
&+K_UT_U + K_CT_C + K_ST_S.
\label{equation_full_receiver_temperature}
\end{align}

\noindent Here, $P_L$ and $P_{L+NS}$ are the PSDs from the internal ambient load and ambient plus noise source, respectively; $T^a_L$ and $T^a_{L+NS}$ are assumed values for the noise temperatures of the internal ambient load and ambient plus noise source, respectively; $C_1$ is an absolute multiplicative correction to the difference $T^a_{L+NS} - T^a_{L}$; $C_2$ is an absolute additive correction to $T^a_L$; and $T_U$, $T_C$, and $T_S$ are the noise wave parameters of the receiver front-end in the formalism of \citet{meys1978}. $C_1$, $C_2$, $T_U$, $T_C$, and $T_S$ are referred to as the absolute receiver calibration parameters, which we determine using the measurements from the external calibration standards \citep{monsalve2017a}. The $K$ parameters in Equations~\ref{equation_full_receiver_gain} and \ref{equation_full_receiver_temperature} capture the impedance mismatch between the antenna and the receiver input, and are given by:

\begin{align}
K_0&=\frac{1-|\Gamma_R|^2}{\left(1-|\Gamma_A|^2\right)|F|^2}, \label{equation_K0}\\
K_U&=\frac{|\Gamma_A|^2}{1-|\Gamma_A|^2}, \\
K_C&=\frac{|\Gamma_A|}{\left(1-|\Gamma_A|^2\right)|F|} \cos\alpha, \\
K_S&=\frac{|\Gamma_A|}{\left(1-|\Gamma_A|^2\right)|F|} \sin\alpha, \\
F&=\frac{\sqrt{1-|\Gamma_R|^2}}{1-\Gamma_A\Gamma_R}, \\
\alpha&=\mathrm{arg}(\Gamma_{A}F), \label{equation_alpha}
\end{align}

\noindent where $\Gamma_A$ is the reflection coefficient of the antenna, including the effect of the balun, and $\Gamma_R$ is the reflection coefficient looking into the receiver input.

\section{Antenna}
\label{section_antenna}
Similarly to other experiments \citep[e.g.][]{anstey2022,cummer2022}, in MIST we explored a variety of antenna types. In addition to having a zenith-pointing beam (Section~\ref{section_description}), the antenna must produce a good impedance match with the receiver (Section~\ref{section_formalism_receiver}) and have a geometry that is easy to simulate electromagnetically. After considering the tradeoffs between these criteria, we selected a horizontal blade dipole antenna for MIST. 

The blade dipole antenna was introduced for global $21$~cm measurements by EDGES-2 in \citet{mozdzen2016}.\footnote{In its first iteration, EDGES used a four-point antenna.} However, in MIST we use this antenna without a ground plane, which represents a significant difference. Moreover, in MIST all the dimensions of the antenna and instrument (Table~\ref{table_antenna_dimensions}) are different from those in EDGES. These differences offer a valuable opportunity for the cross-checking of systematics between MIST and EDGES. The differences between MIST and other current experiments are even more substantial. In particular, SARAS3 uses a vertical monopole over water \citep{singh2022} and REACH uses two antenna types: a hexagonal dipole and a conical log-spiral, both above a $20$~m $\times$~$20$~m metal ground plane that is suspended above the soil \citep{deleraacedo2022}. With its unique characteristics, the MIST antenna represents a significant contribution to the experimental parameter space of global $21$~cm instruments.

\begin{table}
\caption{Dimensions of the MIST instrument. The parameters correspond to those in the schematic of Figure~\ref{figure_instrument_diagram}.}             % title of Table
\label{table_antenna_dimensions}      % is used to refer this table in the text
\centering                          % used for centering table
\begin{tabular}{l c c}        % centered columns (4 columns)
\hline % inserts double horizontal lines
\\
Dimension & Parameter & Value (m) \\ % table heading 
\hline                        % inserts single horizontal line
Antenna panel length       & al &  $1.2$   \\
Antenna panel width        & aw &  $0.6$   \\ 
Antenna panel thickness    & at &  $0.003$ \\ 
Antenna panel separation   & as &  $0.02$  \\
Antenna panel height       & ah &  $0.52$   \\ 
Receiver length         & rl &  $0.405$ \\ 
Receiver width          & rw &  $0.26$ \\ 
Receiver depth          & rd &  $0.335$  \\
Receiver height         & rh &  $0.02$  \\
Balun length         & bl &  $0.05$  \\
Balun width         & bw &  $0.03$  \\
Balun depth         & bd &  $0.037$  \\
Balun height         & bh &  $0.022$  \\
\hline                                   %inserts single line
\end{tabular}
\end{table}

\begin{figure}
\centering
\includegraphics[width=\linewidth]{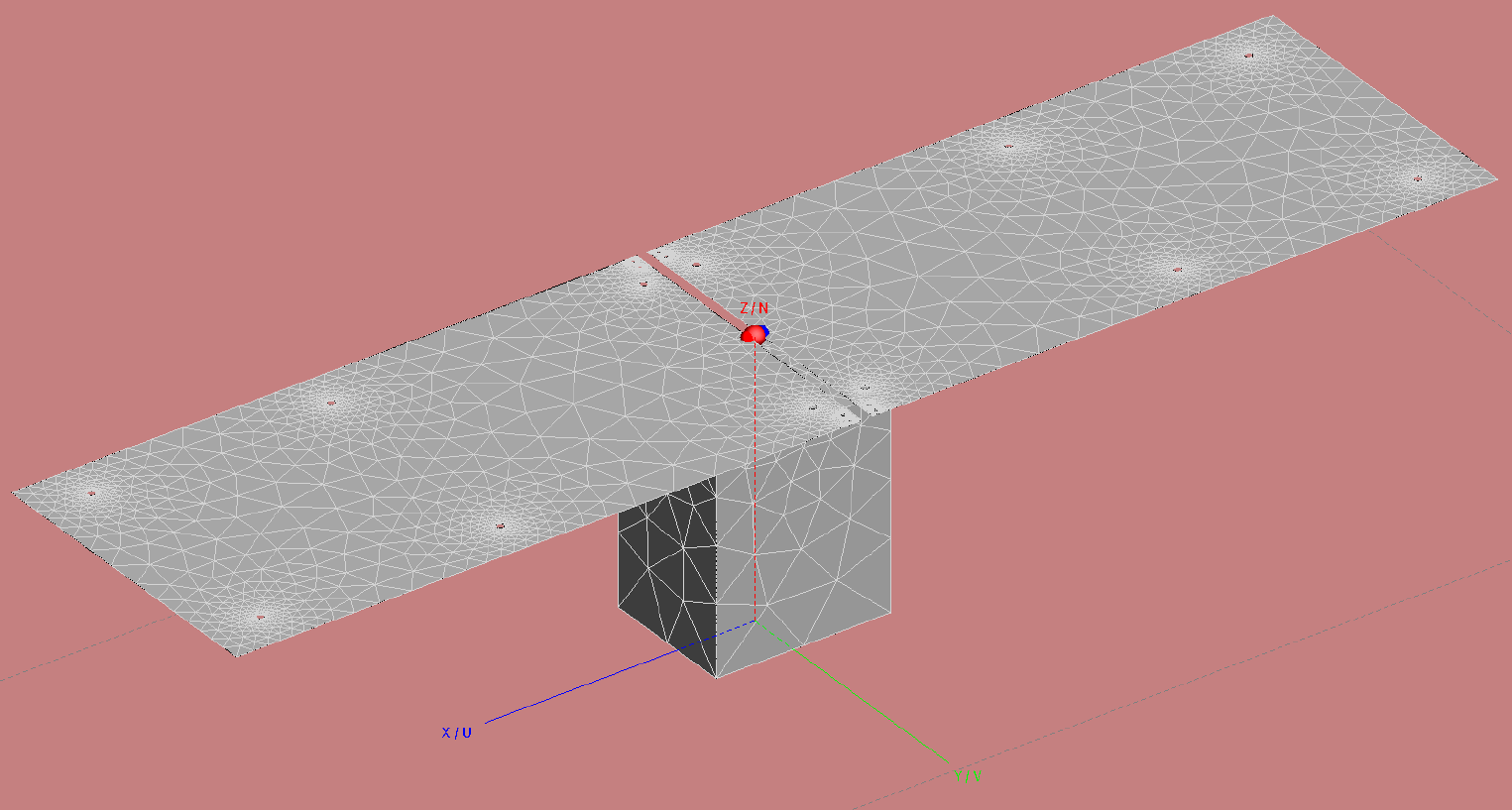}
\caption{Geometry of the MIST simulations in the FEKO software. The geometry includes the antenna panels, balun (not seen in the figure because it is under the antenna panels), and receiver box. The dimensions of the instrument in the simulations are those shown in Table~\ref{table_antenna_dimensions}. The excitation port is simulated as a voltage source in the middle of a wire connecting the two antenna panels. The meshing seen in the figure corresponds to a triangular discretization done on the two-dimensional surfaces of the geometry for the computation of currents and fields with the method of moments. The geometry also includes soil under the instrument, which extends to infinity in the horizontal direction and depth. The antenna support frame, made out of fiberglass, nylon, and ABS plastic, does not have a significant effect on the antenna response and is not included in the simulations.}
\label{figure_FEKO}
\end{figure}

\subsection{Electromagnetic simulations}
\label{section_feko_simulations}

We use electromagnetic simulations with the FEKO software\footnote{\url{https://www.altair.com/feko}} to predict the parameters of the MIST antenna necessary for the calibration of the sky measurements ---radiation efficiency, beam directivity, beam efficiency, and reflection coefficient (Section~\ref{section_formalism}). In FEKO, we use the method of moments solver, which is the most suited for radiation problems involving electrically large structures.

\subsubsection{Simulation geometry}

The simulation geometry includes the antenna panels, balun, and receiver box, following the schematic of Figure~\ref{figure_instrument_diagram} and dimensions of Table~\ref{table_antenna_dimensions}. The geometry also includes soil, which extends to infinity in the horizontal direction and depth.\footnote{The effect of the soil is computed analytically using Sommerfeld integrals as Green's functions for solving the boundary conditions \citep{davidson2011,mosig2021}.} Figure~\ref{figure_FEKO} shows a screenshot of the simulations. The antenna panels are modeled as $3$-mm thick aluminum sheets. Each panel includes the holes illustrated in Figure~\ref{figure_instrument_diagram}: six of these holes connect to the support frame, and the four holes near the edge are used to fine-tune the panel-to-panel separation distance. The excitation port is simulated as a voltage source in the middle of a wire connecting the two antenna panels. The diameter of the wire is $1$~mm, which matches the diameter of the real wires that connect the antenna to the balun (Section~\ref{section_balun}). The simulated wire is modeled as a perfect electric conductor because the effect of the finite conductivity of the real wires is included in the S-parameters of the balun. The balun and receiver box are modeled as hollow boxes with $1.6$-mm thick aluminum walls, matching the real instrument. We use an electrical conductivity of $3.816\times10^{7}$~Sm$^{-1}$ for aluminum \citep{gardiol1984}. The antenna support frame (made out of fiberglass, nylon, and ABS plastic) does not have a significant effect on the antenna response and is not included in the simulations.

\subsubsection{Soil models}

The performance of the antenna depends very strongly on the electrical parameters of the soil: specifically, the conductivity ($\sigma$) and relative permittivity ($\epsilon_r$). In real soils, $\sigma$ and $\epsilon_r$ are spatially inhomogeneous. However, in FEKO it is only possible to design infinite soil models with variations as a function of depth. These variations are implemented using horizontal layers of different $\sigma$ and $\epsilon_r$ \citep{spinelli2022}. To simplify the modeling and work within the capabilities of FEKO, for MIST we must choose observation sites where the soil is very close to flat and any variations primarily occur as a function of depth.

To explore the effect of soil on the MIST antenna performance, we run FEKO simulations with nine models for the soil. The characteristics of the models are listed in Table~\ref{table_soil_parameters}. Five of these models, labelled `nominal' and \verb+1L_xx+, are single-layer models. These models intend to mimic the optimistic scenario in which, from the point of view of the antenna, the soil can be well described as effectively uniform. The nominal model, which in this paper is used as a reference, has values $\sigma_1=0.01$~Sm$^{-1}$ and $\epsilon_{r1}=6$. In the \verb+1L_xx+ models, the value of one of the parameters is changed relative to the nominal model, with the \verb~xx~ part of the label being used to identify the change. Specifically, \verb~c+~ (\verb~c-~) corresponds to an increase (decrease) in conductivity to $0.1$ ($0.001$)~Sm$^{-1}$, and \verb~p+~ (\verb~p-~) represents an increase (decrease) in relative permittivity to $10$ ($2$). The values we use for $\sigma$ and $\epsilon_r$ fall within the ranges reported for several geological materials that we could encounter at our observation sites. For instance, snow and freshwater ice typically have $0\lesssim\sigma\lesssim0.01$~Sm$^{-1}$ and $2\lesssim\epsilon_r\lesssim6$. For sand, silt, and clay, $\sigma$ and $\epsilon_r$ strongly depend on moisture level and span the wide ranges $10^{-7}$--$10^{0}$~Sm$^{-1}$ and $2$--$40$, respectively \citep{reynolds2011}. Our choices for the parameter values, in particular, closely match the values reported by \citet{sutinjo2015} for the soil at the Inyarrimanha Ilgari Bundara, the CSIRO Murchison Radio-astronomy Observatory, with different moisture levels. Our values are also consistent with the soil measurements done by \citet{spinelli2022} at the Owens Valley Radio Observatory for dry and wet conditions.

As shown in Table~\ref{table_soil_parameters}, our last four soil models, labelled \verb+2L_xx+, are two-layer models. The purpose of these models is to represent the more realistic situation in which there is a change in the soil parameters at some depth from the surface. In these models, the top layer has a thickness $L=1$~m and the same conductivity and permittivity as the nominal case, while the bottom layer extends to infinite depth and has a different value of either conductivity or permittivity. In each case, the parameter change made to the bottom layer is indicated in the model name, similarly to the single-layer models. Furthermore, the conductivity and relative permittivity values assigned to the bottom layer in the \verb~c+~, \verb~c-~, \verb~p+~, and \verb~p-~ cases are the same as for the single-layer models. As an example, one observation site where the soil could be well represented by a two-layer model is MARS during the summer. At that site, the soil in the summer consists of an unfrozen top layer and a permanently frozen, or `permafrost', bottom layer \citep[e.g.][]{pollard2009,wilhelm2011}. We leave for future work the exploration of a wider range of soil models, including models with more than two layers (which were studied in \citet{spinelli2022} for the LEDA experiment), and models that account for the fact that in real soils the conductivity and permittivity vary as a function of frequency \citep[e.g.][]{revil2013}.

\subsubsection{Simulation settings}

The simulations are conducted in the range $25$--$125$~MHz with a resolution of $1$~MHz. Although the sky measurements are currently analysed up to $105$~MHz, the simulations extend up to $125$~MHz anticipating a future increase in the bandwidth of MIST. In FEKO, we use the \verb+fine+ mesh size setting and double-precision calculations. Using the $64$ cores of an AMD Ryzen Threadripper PRO 5995WX processor, each simulation takes $\approx60$~minutes.

Next, we describe the results of the FEKO simulations for the radiation efficiency, beam directivity, beam efficiency, and reflection coefficient at the dipole excitation port.

\begin{table}
\caption{Soil models used in the FEKO simulations of MIST. Five of these models are single-layer models (``nominal'' and \texttt{1L\_xx}) and four are two-layer models (\texttt{2L\_xx}). The layers are characterized in terms of their electrical conductivity ($\sigma$) and relative permittivity ($\epsilon_r$). In the two-layer models, the thickness of the top layer is $L=1$~m.}             % title of Table
\label{table_soil_parameters}      % is used to refer this table in the text
\centering                          % used for centering table
\begin{tabular}{l c c c c c}        % centered columns (4 columns)
\hline % inserts double horizontal lines
\\
Model & \# layers & $\sigma_1$~[Sm$^{-1}$] & $\epsilon_{r1}$ & $\sigma_2$~[Sm$^{-1}$] & $\epsilon_{r2}$\\ % table heading 
\hline                        % inserts single horizontal line
nominal    & 1 & $0.01$ &  $6$ & & \\
\hline
\verb~1L_c+~       & 1 & $0.1$ &  $6$ & & \\
\verb~1L_c-~       & 1 & $0.001$ &  $6$ & &\\
\verb~1L_p+~       & 1 & $0.01$ &  $10$ & & \\
\verb~1L_p-~       & 1 & $0.01$ &  $2$ & & \\
\hline
\verb~2L_c+~       & 2 & $0.01$ &  $6$ & $0.1$ &  $6$ \\
\verb~2L_c-~       & 2 & $0.01$ &  $6$ & $0.001$ &  $6$ \\
\verb~2L_p+~       & 2 & $0.01$ &  $6$ & $0.01$ &  $10$ \\
\verb~2L_p-~       & 2 & $0.01$ &  $6$ & $0.01$ &  $2$ \\
\hline                                   %inserts single line
\end{tabular}
\end{table}

\begin{figure*}
\centering
\includegraphics[width=\linewidth]{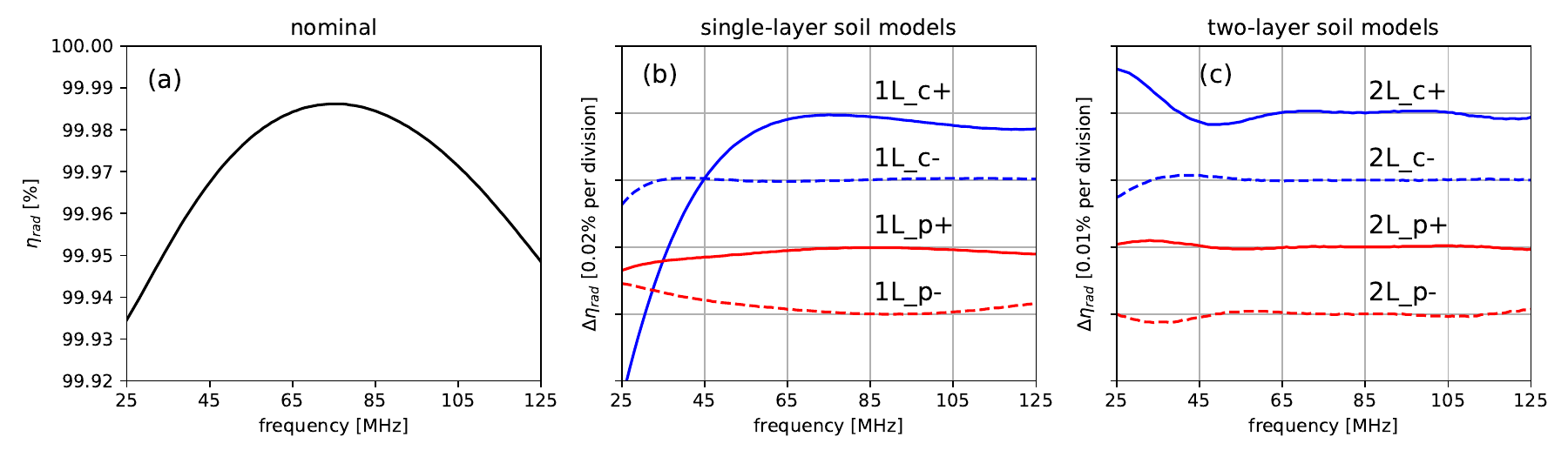}
\caption{(a) Simulated radiation efficiency for the nominal soil model. (b) and (c) Differences in radiation efficiency between the alternative soil models and the nominal model. For example, for model \texttt{1L\_c+}, $\Delta\eta_{rad}=\eta_{rad,\;\texttt{1L\_c+}}-\eta_{rad,\;nominal}$. In panels (b) and (c), the zero-points for the differences are the labelled horizontal grid lines.}
\label{figure_radiation_efficiency}
\end{figure*}

\subsection{Radiation efficiency}
The radiation efficiency of the antenna, $\eta_{rad}$, is defined as the ratio of radiated power to input power. $\eta_{rad}$ relates the beam gain, $G$, to the beam directivity by

\begin{align}
G(\theta,\phi,\nu) = \eta_{rad}(\nu) D(\theta,\phi,\nu).
\end{align}

\noindent In this definition, the efficiency only considers resistive, or `Ohmic', loss in the antenna and conductive regions visible to the antenna, and not impedance mismatch or external losses, such as ground loss \citep{stutzman1998}. In our FEKO simulations, we assign the finite conductivity of aluminum to the antenna panels, balun, and receiver box, and the conductivities of Table~\ref{table_soil_parameters} to the soil. This information enables FEKO to calculate and directly provide $\eta_{rad}$ and $G$. The beam directivity is a derived quantity and obtained from $\eta_{rad}$ and $G$ (Section~\ref{section_beam_directivity}).

Figure~\ref{figure_radiation_efficiency} shows the radiation efficiency for the nine FEKO simulations. Panel~(a) shows that for the nominal soil model, the efficiency is in the range $99.93\%$--$99.99\%$ and has a smooth frequency dependence. This high efficiency is expected for the high conductivity of the instrument's aluminum surfaces. Panels~(b) and (c) show the difference in the radiation efficiency for the alternative soil models relative to the nominal model. The main takeaway of these panels is the confirmation that the radiation efficiency is sensitive to the characteristics of the soil, in addition to those of the instrument's surfaces. For our single-layer (two-layer) models, the largest change is $\approx0.08\%$ ($\approx0.007\%$). Although small compared to the total efficiency, these changes are comparable to, or larger than, the ratio between the global $21$~cm signal and the diffuse astrophysical foreground, which is $\approx0.01\%$ for the Cosmic Dawn and smaller for the Dark Ages. This comparison highlights the need to accurately determine and account for the radiation efficiency in order to minimize biases in the $21$~cm signal estimation.

\begin{figure}
\centering
\includegraphics[width=\linewidth]{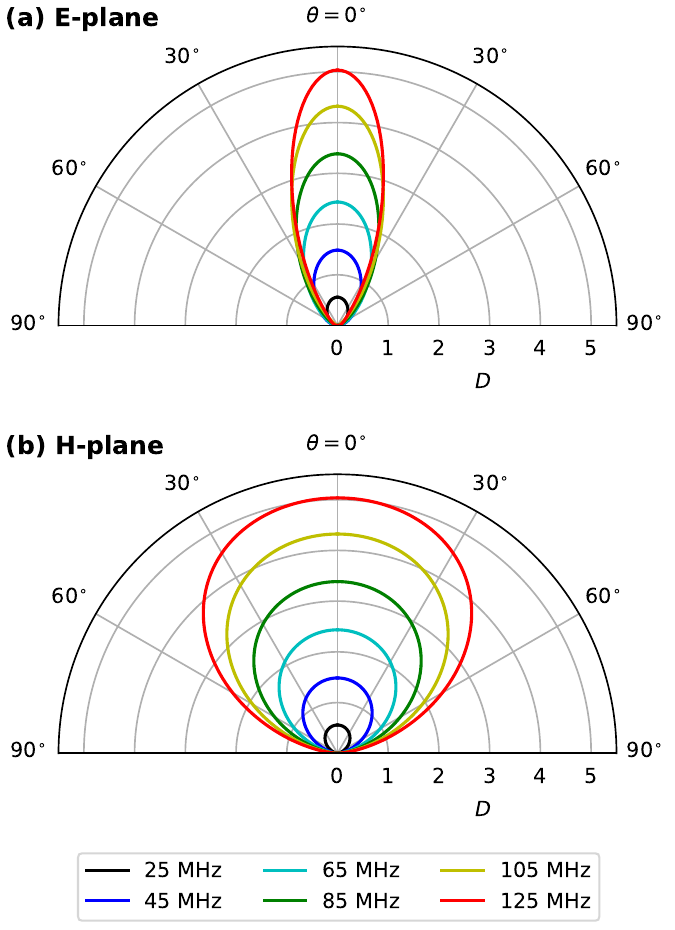}
\caption{Simulated antenna beam directivity for the nominal soil model in the E- and H- planes at six frequencies.}
\label{figure_beam_directivity1}
\end{figure}

\subsection{Beam directivity}
\label{section_beam_directivity}

We compute the beam directivity as $D=\eta_{rad}^{-1}$G using the gain and radiation efficiency provided by FEKO. The directivity is only computed for the top hemisphere because FEKO cannot provide the gain in the soil when the soil has non-zero electrical conductivity. In MIST, the effect of the directivity in the soil is accounted for through the beam efficiency, which is calculated using the top-hemisphere directivity (Section~\ref{section_ground_loss}).

Figure~\ref{figure_beam_directivity1} shows slices of the directivity every $20$~MHz for the nominal soil model. The slices are shown for the E- and H- planes, which are parallel and perpendicular to the antenna excitation axis (i.e. the antenna's length), respectively. The directivity peaks at the zenith and is minimized at the horizon, satisfying one of the main design objectives for the antenna (Section~\ref{section_description}). The peak directivity increases monotonically from $\approx0.6$ to $\approx5.1$ between $25$ and $125$~MHz, and the directivity pattern is wider in the H-plane at all frequencies.

\begin{figure*}
\centering
\includegraphics[width=\linewidth]{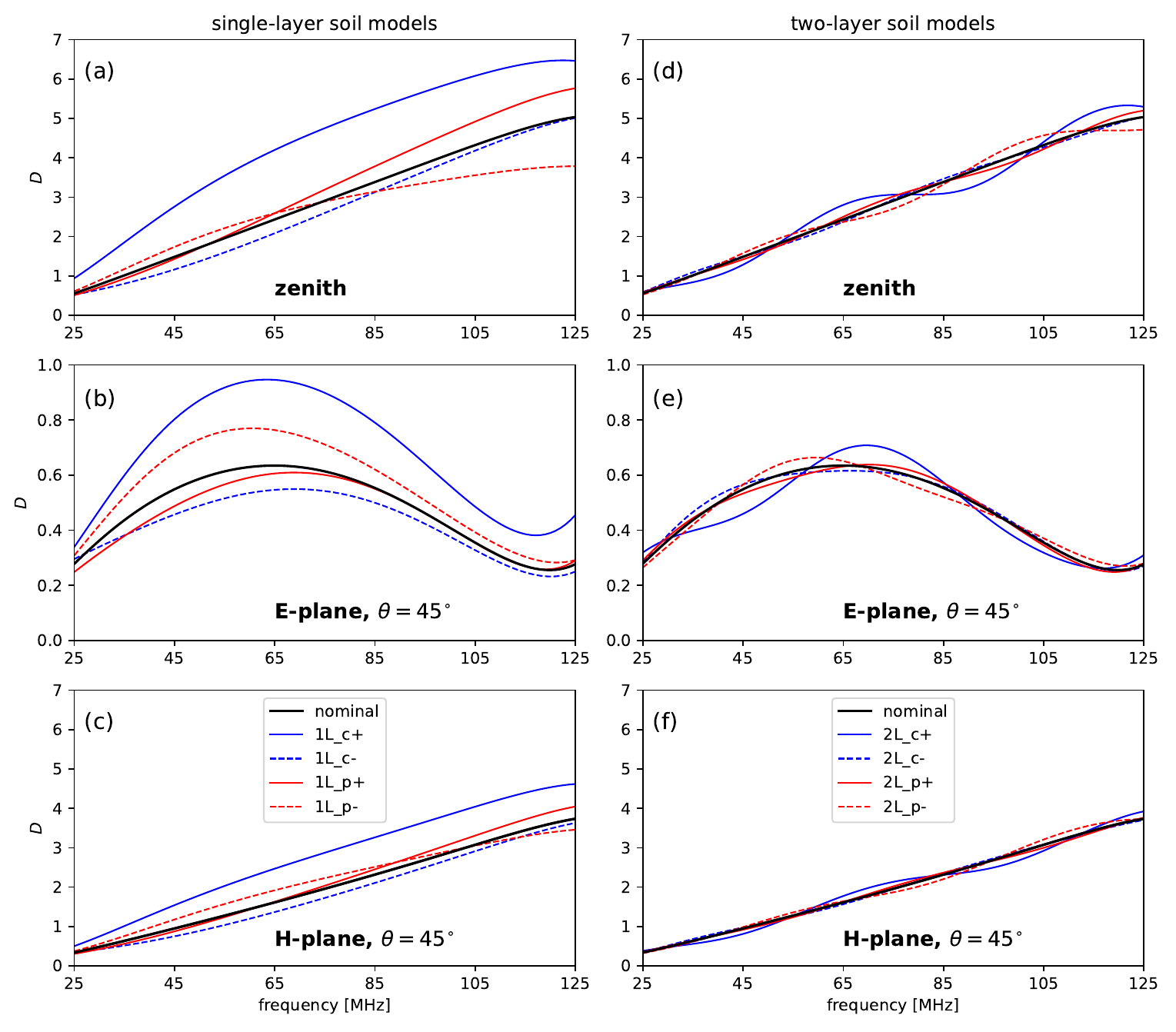}
\caption{Simulated antenna beam directivity for all the soil models at three reference puncture points: (\textit{top row}) zenith; (\textit{middle row}) E-plane at $\theta=45^{\circ}$; and (\textit{bottom row}) H-plane at $\theta=45^{\circ}$.}
\label{figure_beam_directivity5}
\end{figure*}

\begin{figure*}
\centering
\includegraphics[width=\linewidth]{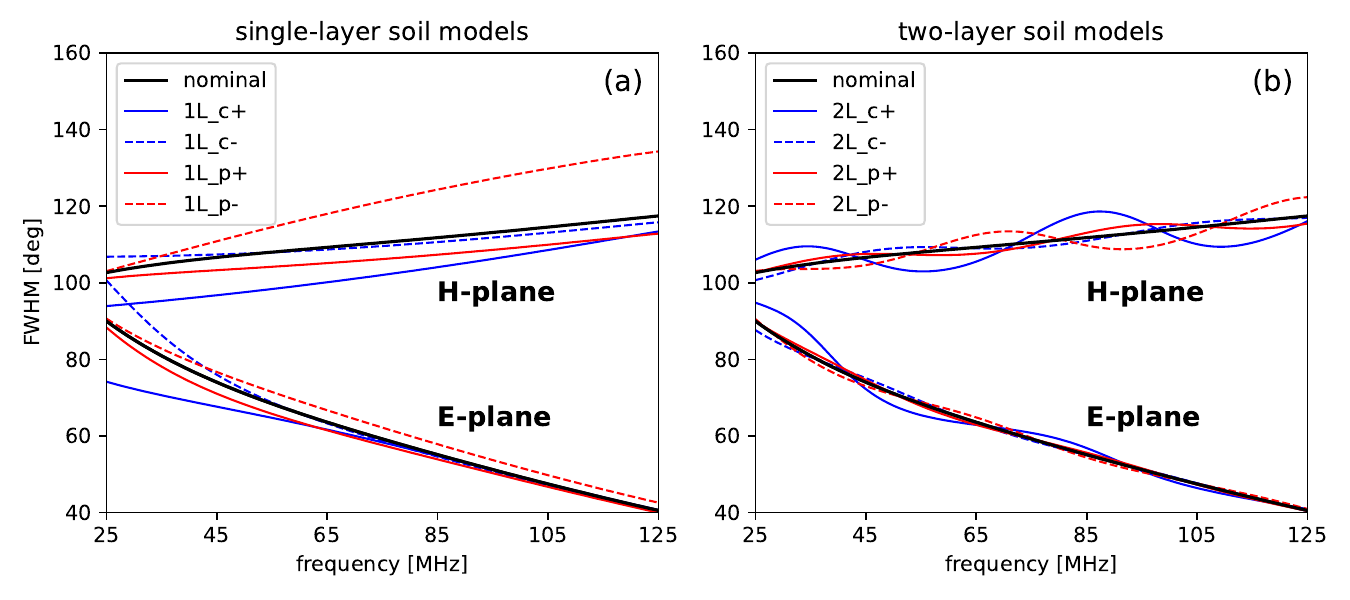}
\caption{FWHM of the simulated antenna beam directivity in the E- and H-planes for all the soil models.}
\label{figure_beam_directivity4}
\end{figure*}

Figure~\ref{figure_beam_directivity5} shows the directivity for the nine soil models at three reference puncture points: (1) the zenith, (2) the E-plane at $\theta=45^{\circ}$, and (3) the H-plane at $\theta=45^{\circ}$. In this figure, we identify the following important points: 

\begin{enumerate}[(1)]
\item The directivity is very sensitive to the properties of the soil. As expected, the effects are larger when the changes in the soil extend all the way to the surface, which occurs in the single-layer models. For these models, the directivity increases with conductivity across most of the frequency range. In particular, the directivity for model \verb~1L_c+~ is $>20\%$, and at some frequencies $>50\%$, higher than the nominal. Changes in permittivity also produce significant changes in the directivity, but the sign of the changes evolves with frequency.

\item The uniformity of the single-layer soil models results in a directivity that is relatively smooth across frequency. For two-layer models, changes in the bottom layer introduce ripples which, for the $1$-m thickness of the top layer, have a period of $\approx50$~MHz.

\item At the zenith and the H-plane puncture point, the directivity increases with frequency for all soil models to values that, at the highest frequencies, are $>3$. At the E-plane puncture point, the directivity is always $<1$ and peaks at $\approx55$--$75$~MHz. The directivities at the E- and H-plane puncture points indicate that the beam is wider in the H-plane for all the soil models.

\end{enumerate}

As a direct quantification of the beam width, Figure~\ref{figure_beam_directivity4} shows the FWHM of the directivity pattern in the E- and H-planes for the nine soil models. For all the models, the FWHM is larger in the H-plane than in the E-plane. Moreover, in the H-plane the FHWM increases with frequency while in the E-plane it decreases. For our soil models, the H-plane FWHM is in the range $85^{\circ}$--$107^{\circ}$ at $25$~MHz, increasing to $113^{\circ}$--$135^{\circ}$ at $125$~MHz. In the E-plane, the FWHM is in the range $74^{\circ}$--$101^{\circ}$ at $25$~MHz, decreasing to $39^{\circ}$--$43^{\circ}$ at $125$~MHz. In both planes, the FWHM increases with decreasing permittivity at the soil surface, as seen when comparing single-layer cases \verb~1L_p+~ and \verb~1L_p-~ with the nominal. Similarly to the directivity (Figure~\ref{figure_beam_directivity5}), the frequency evolution of the FWHM for the single-layer models is relatively smooth. For the two-layer models, the FWHM shows ripples with a period of $\approx50$~MHz and a peak amplitude of up to $\approx8^{\circ}$ with respect to the nominal model. The amplitude of the ripples is larger in the H-plane than in the E-plane.  

The beam directivity of a global $21$~cm instrument has a very strong impact on the accuracy of the signal recovery \citep[e.g.][]{mahesh2021,anstey2022,cummer2022,spinelli2022}. We discuss the $21$~cm signal extraction accuracy expected for the MIST beam directivity and different soil characteristics in \citet{monsalve2024}.

\begin{figure*}
\centering
\includegraphics[width=\linewidth]{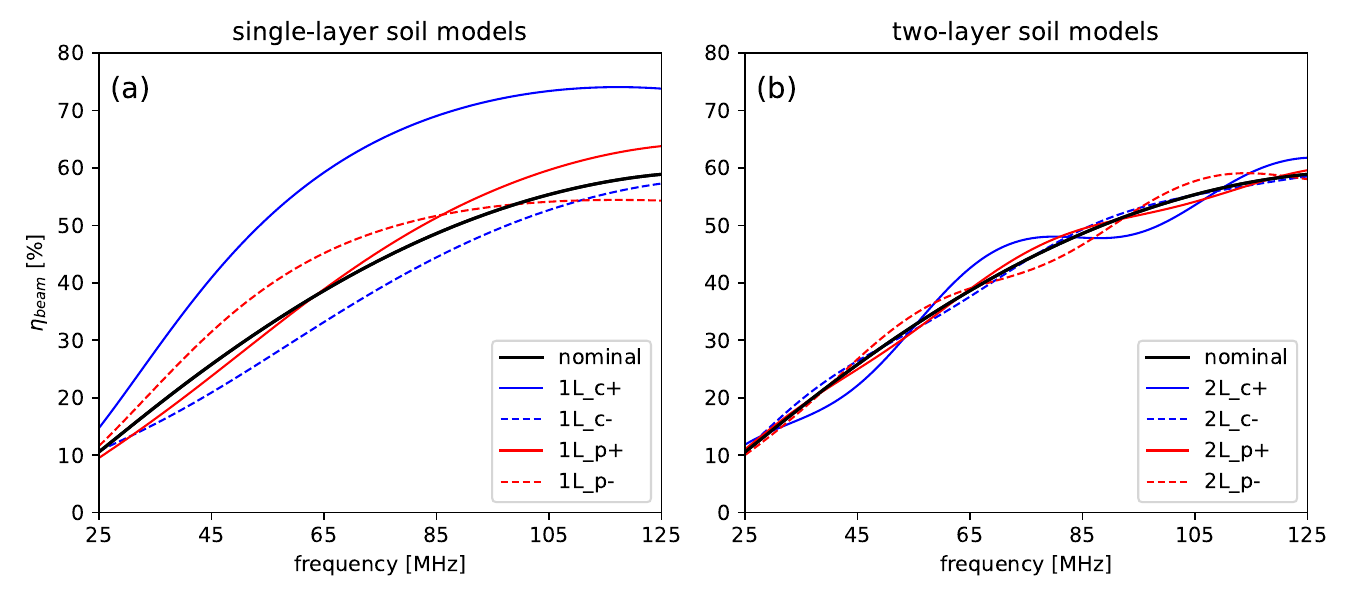}
\caption{Simulated beam efficiency for all the soil models.}
\label{figure_beam_efficiency}
\end{figure*}

\subsection{Beam efficiency}
\label{section_ground_loss}

We define the beam efficiency, $\eta_{beam}$, as the solid angle of the beam directivity in the top hemisphere divided by the solid angle over the full sphere, i.e.

\begin{align}
\eta_{beam}(\nu) = \frac{1}{4\pi}\int_0^{2\pi} \int_{0}^{\pi/2} D(\theta,\phi,\nu) \sin\theta d\theta d\phi.
\end{align}

This efficiency is equivalent to one minus the ground loss fraction. Figure~\ref{figure_beam_efficiency} shows the beam efficiency for the nine soil models. The main trend observed across models is an increase of the efficiency with frequency. The efficiency is in the range $9\%$--$15\%$ at $25$~MHz, increasing to $54\%$--$74\%$ at $125$~MHz. The efficiency also increases with the conductivity at the soil surface, as seen when comparing single-layer cases \verb~1L_c+~ and \verb~1L_c-~ with the nominal. This dependence is consistent with intuition: The efficiency would be $100\%$ if the soil surface had infinite conductivity, or if an infinitely large and infinitely conducting ground plane were used. As expected from the directivity and FWHM plots (Figures~\ref{figure_beam_directivity5} and \ref{figure_beam_directivity4}), the beam efficiency has a relatively smooth frequency evolution for single-layer models, while for two-layer models it contains ripple-like structure. This structure has a period of $\approx50$~MHz and a peak amplitude of up to $\approx5\%$ with respect to the nominal model.

\begin{figure*}
\centering
\includegraphics[width=1\linewidth]{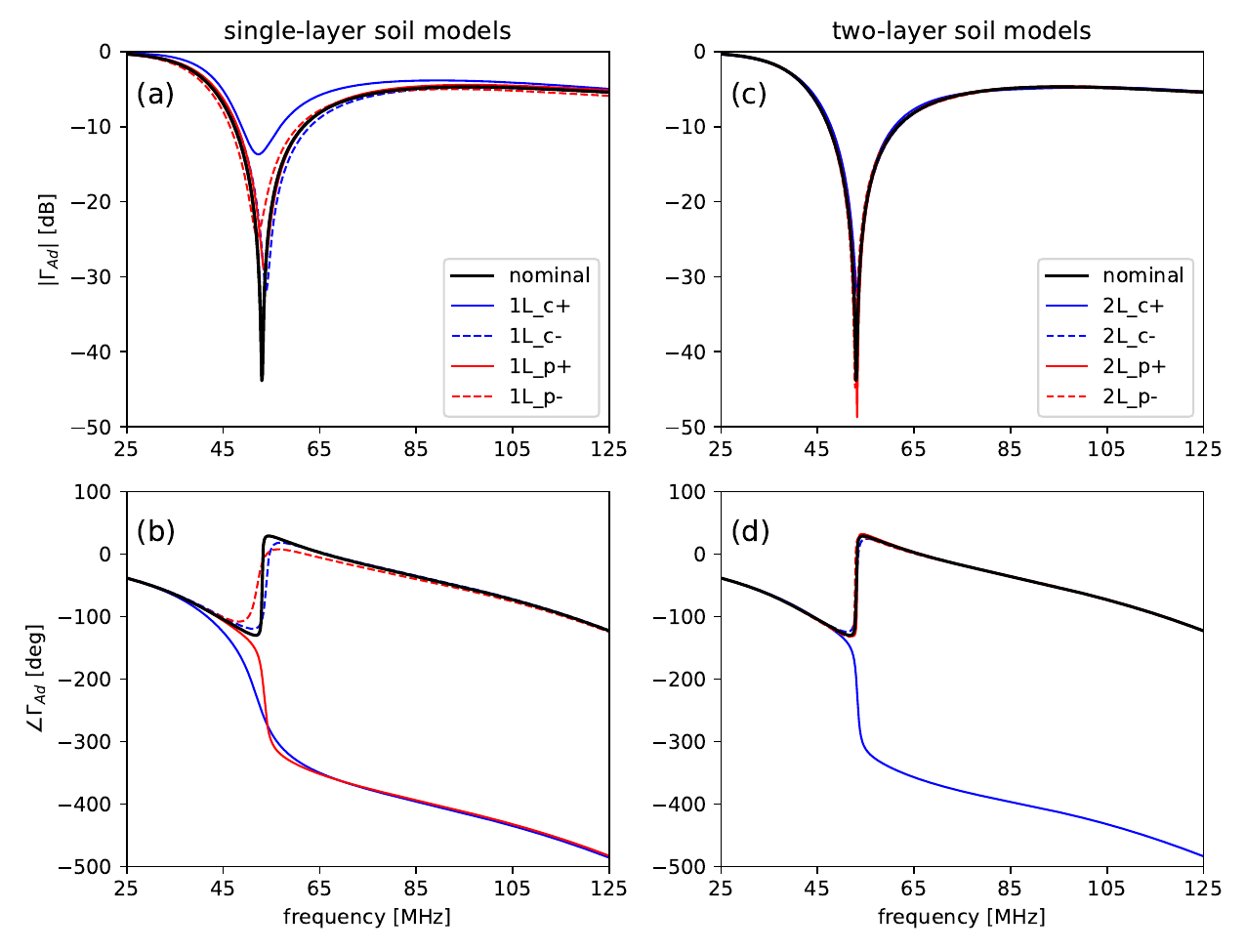}
\caption{Simulated reflection coefficient at the dipole excitation port for all the soil models.}
\label{figure_impedance_port}
\end{figure*}

\subsection{Reflection coefficient at dipole excitation port}
\label{section_antenna_impedance}

Our FEKO simulations provide the reflection coefficient of the antenna at the dipole excitation port, $\Gamma_{Ad}$. The reflection coefficient at the balun output, $\Gamma_A$, which is the quantity required for receiver calibration (Section~\ref{section_formalism_receiver}), is described in Section~\ref{section_balun_reflection_coefficient}. 

Figure~\ref{figure_impedance_port} shows the reflection coefficient at the dipole excitation port. The main characteristic of the reflection magnitude (top row) is a resonant feature at $\approx52$~MHz, where the magnitude reaches a minimum. From this resonance, the magnitude increases monotonically toward lower frequencies until it reaches $\approx0$~dB at $25$~MHz. The magnitude also increases toward higher frequencies and has a peak of $\approx-4$~dB at $90$--$95$~MHz. For single-layer models, the reflection magnitude shows a strong dependence on soil parameters. The largest effect is at the resonance, where the magnitude varies from $\approx-15$ to $-45$~dB across our models. The soil parameters also affect the resonance frequency. Specifically, this frequency decreases with conductivity and increases with permittivity. For our models, these changes are within a few MHz. For two-layer models, changes in the bottom layer produce changes in reflection magnitude that are typically within $2$~dB, except at the resonance.

The bottom row of Figure~\ref{figure_impedance_port} shows the reflection phase. The phase has a predominantly descending slope and a relatively fast $\approx180^{\circ}$ transition at the resonance. Changes in the soil impact the phase primarily around the resonance. For most of our models, when going through the resonance from lower to higher frequencies the phase increases. However, for some models the phase decreases, leading to a $\approx360^{\circ}$ difference across models above the resonance. The negative slope of the phase corresponds to the antenna delay. Except around the resonance, the delay is similar across our models. Specifically, the phase change between $25$ and $45$~MHz (below the resonance) is $\approx60^{\circ}$, corresponding to a delay of $\approx(60^{\circ}/ 360^{\circ}) \times (20 \;\mathrm{MHz})^{-1} \approx 8\;\mathrm{ns}$. Between $55$ and $125$~MHz (above the resonance) the phase change is $\approx150^{\circ}$ and the delay is $\approx6$~ns.

\subsection{Variation of top layer thickness}

\begin{figure}
\centering
\includegraphics[width=1\linewidth]{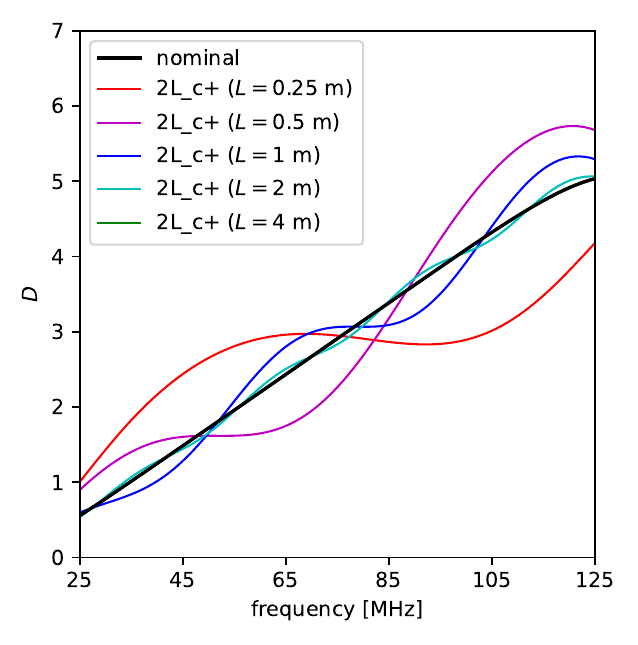}
\caption{Ripples in the simulated antenna beam directivity at the zenith for the \texttt{2L\_c+} two-layer soil model. The only parameter being varied is the thickness of the top layer, $L$. The directivity for the nominal single-layer model is also shown for reference. Relative to the nominal directivity, as $L$ increases, both the amplitude and period of the ripples decrease. The directivity for $L=4$~m is not visible because it is almost perfectly overlapped by the nominal directivity. The dependence of the ripples on $L$ observed in this example is, qualitatively, representative of the other antenna parameters discussed in Section~\ref{section_antenna}.}
\label{figure_change_thickness}
\end{figure}

So far, the results we have shown for two-layer soil models correspond to cases in which the thickness of the top layer is $L=1$~m. Here, we present one example of how the ripples produced by two-layer models change when the top layer thickness is varied in the range $0.25$--$4$~m. For this example, we use the \verb~2L_c+~ model because it is the one that, among our two-layer models with $L=1$~m, produces the largest ripples, making the results easier to see and interpret. In this example, we only show the beam directivity at the zenith, which qualitatively represents the behavior of the ripples in the other antenna parameters. The results are shown in Figure~\ref{figure_change_thickness}. As the top layer thickness increases, the period of the ripples in the directivity decreases. This decrease occurs because of the increased delay of the signal reflected back to the antenna from the interface between the two layers. As $L$ increases, the amplitude of the ripples also decreases. This decrease is due to the increased attenuation suffered by the reflected signal as it travels a larger round-trip distance through the lossy top layer.

\section{Balun}
\label{section_balun}
\subsection{Description}
\label{section_balun_description}

\begin{figure}
\centering
\includegraphics[width=\linewidth]{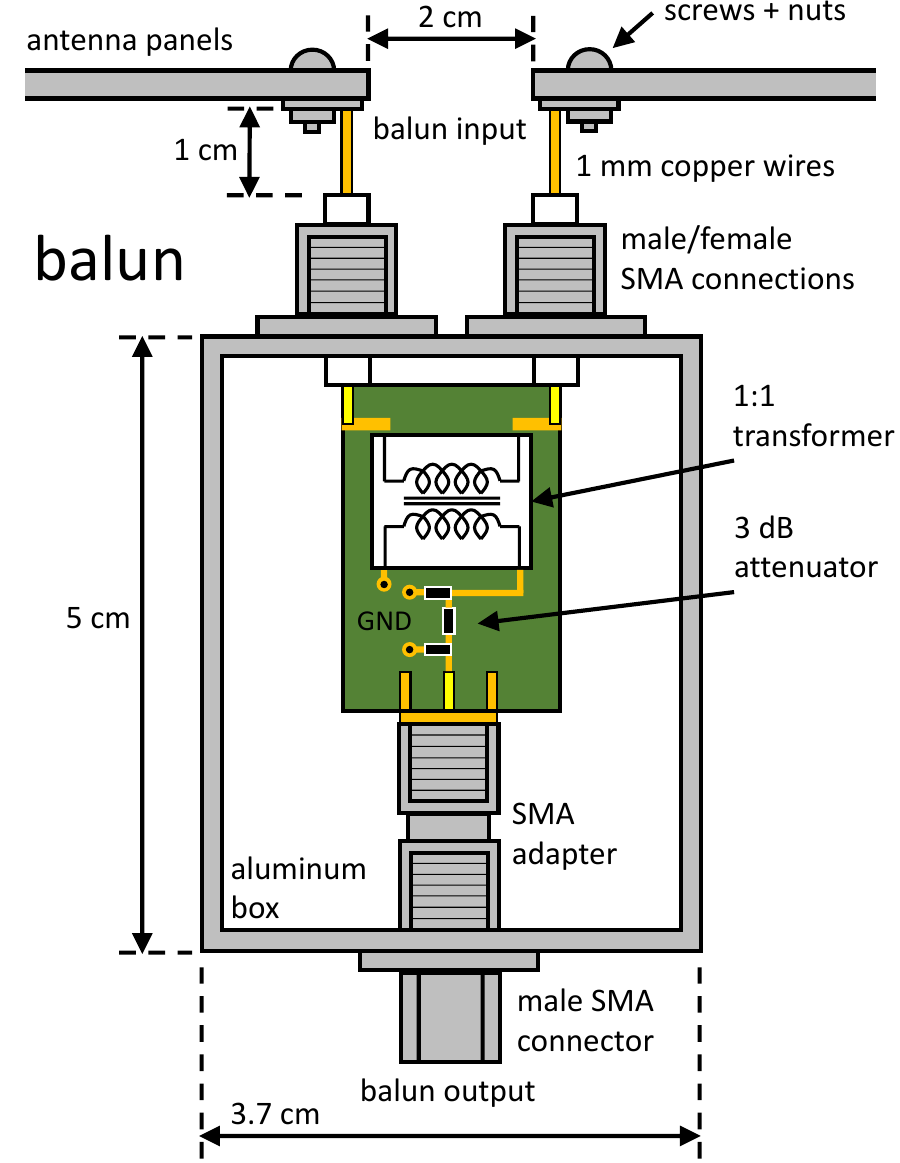}
\caption{Side view schematic of the balun (not to scale). The core components are a 1:1 $50$~$\Omega$ transformer and a $3$~dB attenuator, which are housed inside the aluminum box.}
\label{figure_balun0}
\end{figure}

We use a passive balun to convert the balanced signal produced at the excitation port of the dipole antenna to unbalanced, or ground-referenced, which is how the signal is expected at the receiver input. Figure~\ref{figure_balun0} shows a schematic of the balun. The balun primarily consists of a Mini-Circuits TX1-1+ 1:1 $50$~$\Omega$ transformer in series with a $\approx3$~dB attenuator on the unbalanced side. The transformer provides DC isolation and magnetic coupling between the balanced and unbalanced sides. The attenuator reduces the magnitude of the antenna reflection coefficient seen by the receiver input, which makes the receiver calibration less sensitive to errors in the antenna reflection coefficient measurements \citep{monsalve2017a}. The improved impedance matching from the attenuator comes at a cost of increased signal loss. However, this loss can be accurately measured in the laboratory and subsequently corrected in the sky measurements. To minimize the sensitivity of the balun to temperature fluctuations, the attenuator is constructed with resistors that are temperature-stable to better than $10$~ppm~$^{\circ}$C$^{-1}$. On the balanced side of the balun, i.e. the balun input, the antenna panels are connected to the transformer through two $1$-cm long wires of $1$-mm diameter in series with male and female subminiature version A (SMA) connectors. The unbalanced side of the balun, i.e. the balun output, connects to the receiver input through a male SMA connector.

\begin{figure}
\centering
\includegraphics[width=\linewidth]{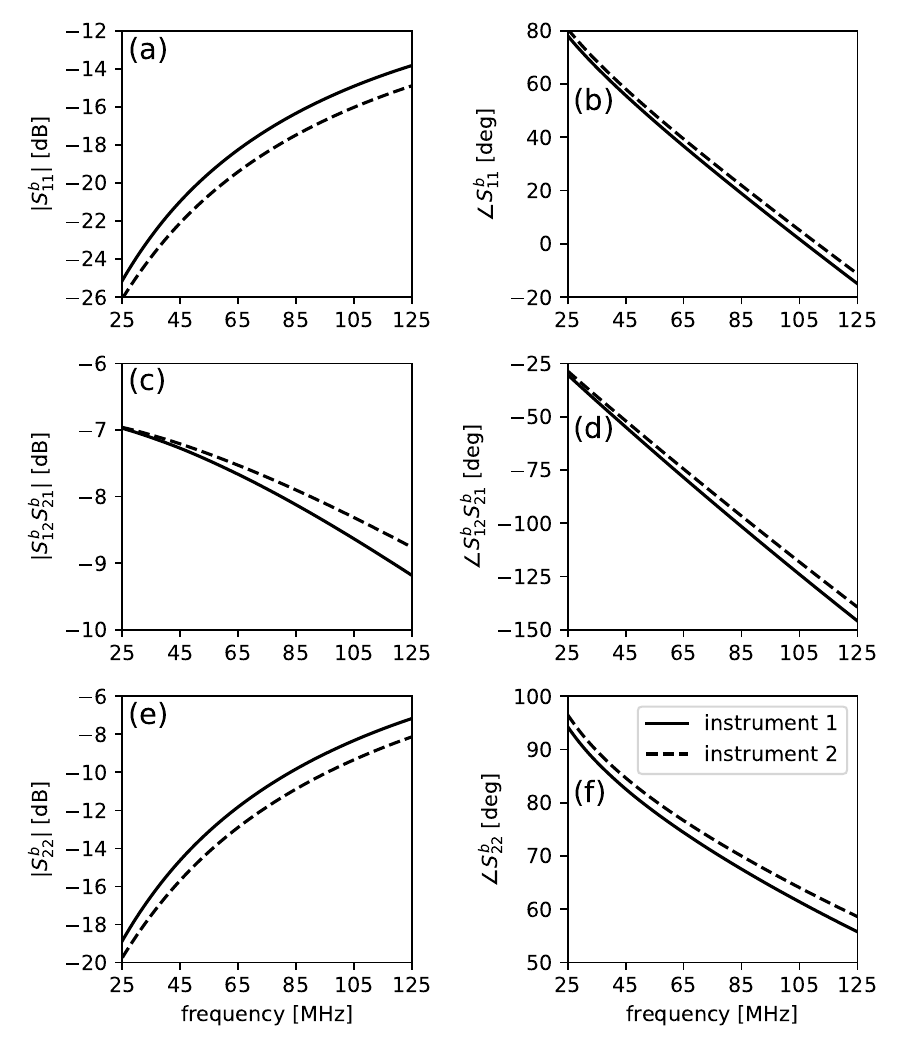}
\caption{S-parameters of the MIST baluns measured in the lab. Here, port~$1$ is the unbalanced port facing the receiver input (balun output) and port~$2$ is the balanced port facing the antenna excitation port (balun input). The parameters include the effect of the wires connecting the antenna panels to the transformer. These measurements were calibrated using a Keysight 85033E $3.5$-mm VNA calibration kit. To maximize the calibration accuracy, for the $50$~$\Omega$ calibration load we used its DC resistance, which was measured with a $5.5$-digit multimeter and the four-wire method, and the delay estimated in \citet{monsalve2016}.}
\label{figure_balun1}
\end{figure}

To characterize the balun, we measured its S-parameters in the lab. The measurements, which include the effect of the wires connecting the antenna panels to the transformer, are shown in Figure~\ref{figure_balun1}. These measurements were calibrated using a Keysight 85033E $3.5$-mm calibration kit.\footnote{\url{https://www.keysight.com/us/en/product/85033E/standard-mechanical-calibration-kit-dc-9-ghz-3-5-mm.html}} To increase the accuracy of this calibration, instead of using the nominal value of $50$~$\Omega$ for the impedance of the calibration load, we used the DC resistance of the load, which was measured with a $5.5$-digit multimeter and the four-wire method. Also, instead of assuming zero delay for the load's transmission line, we used the delay estimated in \citet{monsalve2016}.

\begin{figure*}
\centering
\includegraphics[width=1\linewidth]{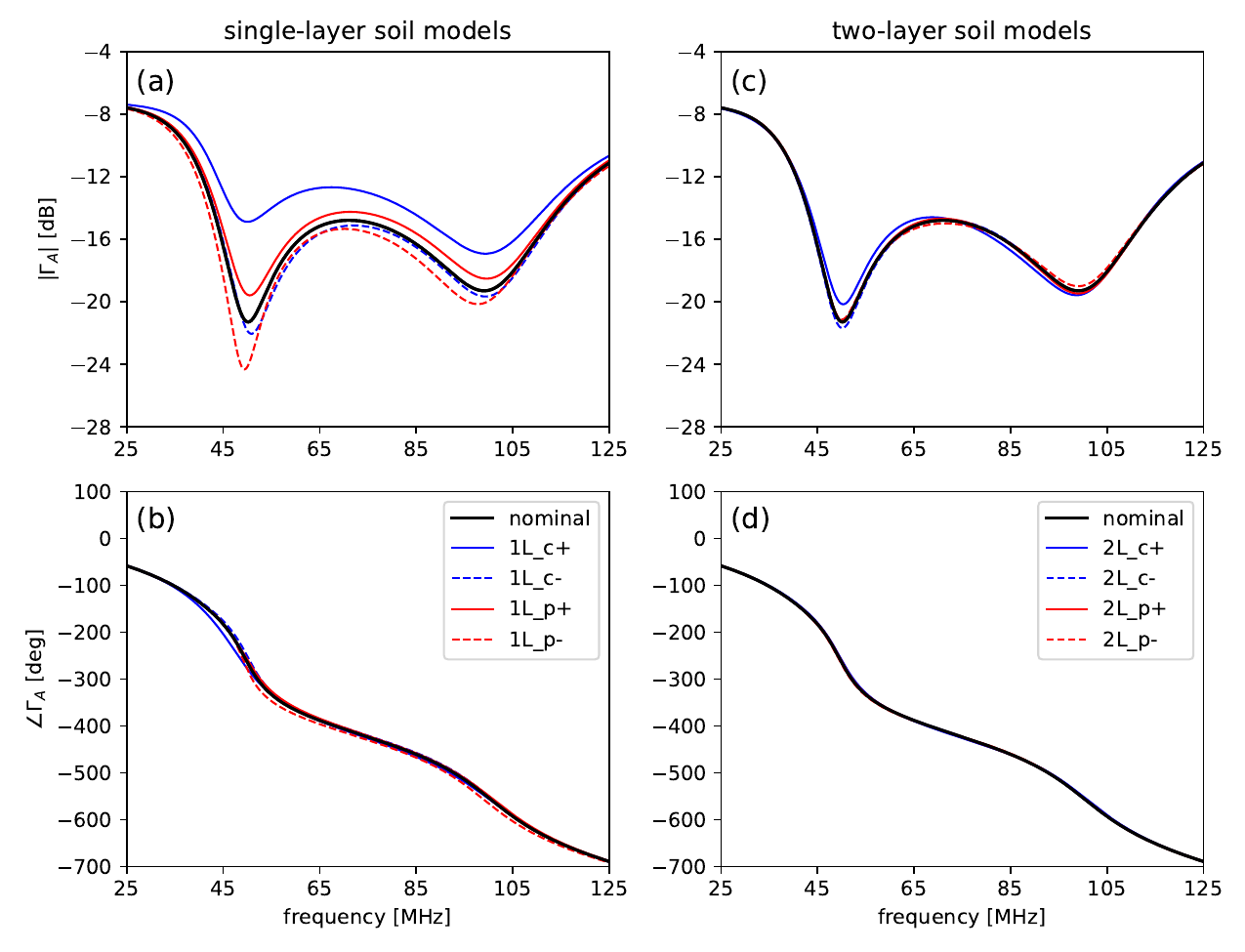}
\caption{Reflection coefficient of the antenna at the balun output for all the soil models. Specifically, the plots show the simulated reflection coefficients from Figure~\ref{figure_impedance_port} projected to the balun output by embedding the measured balun S-parameters from instrument~$1$ (Figure~\ref{figure_balun1}).}
\label{figure_impedance_balun}
\end{figure*}

\begin{figure*}
\centering
\includegraphics[width=\linewidth]{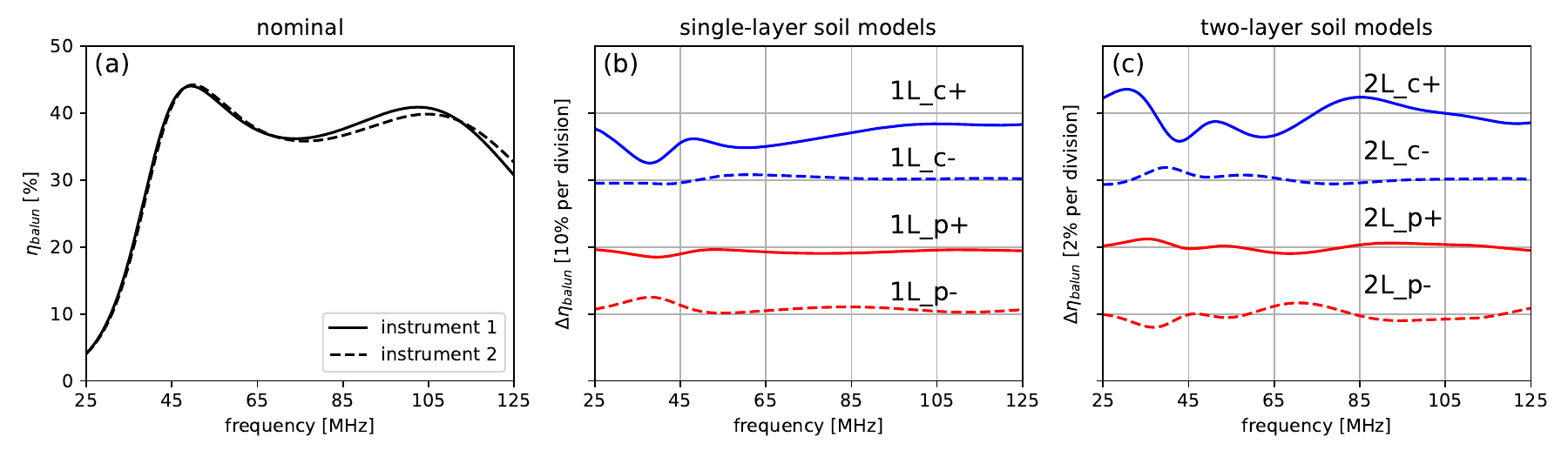}
\caption{(a) Balun efficiency for the two MIST instruments computed with the antenna reflection coefficient for the nominal soil model. (b) and (c) Differences in balun efficiency for instrument~$1$ between the alternative soil models and the nominal model. For example, for model \texttt{1L\_c+}, $\Delta\eta_{balun}=\eta_{balun,\;\texttt{1L\_c+}}-\eta_{balun,\;nominal}$. In panels (b) and (c), the zero-points for the differences are the labelled horizontal grid lines.}
\label{figure_balun3}
\end{figure*}

\subsection{Antenna reflection coefficient at balun output}
\label{section_balun_reflection_coefficient}

To maximize the receiver calibration accuracy, MIST measures the antenna reflection coefficient {\it in situ}. This measurement is done at the balun output. The hardware and techniques used to conduct this measurement are described in Section~\ref{section_receiver}. 

The relationship between the reflection coefficient at the balun output, $\Gamma_{A}$, and at the dipole excitation axis, $\Gamma_{Ad}$, is given by \citep{gonzalez1997}

\begin{equation}
\Gamma_A = S^b_{11} + \frac{S^b_{12}S^b_{21}\Gamma_{Ad}}{1 - S^b_{22}\Gamma_{Ad}}.
\label{equation_gamma_shifted}
\end{equation}

In this equation, $S^b_{11}$, $S^b_{21}$, $S^b_{12}$, and $S^b_{22}$ are the S-parameters of the balun, with port~1~(2) being the balun output (input). The process represented by Equation~\ref{equation_gamma_shifted} is referred to as `embedding' of S-parameters.\footnote{Embedding corresponds to the inclusion of the electrical effects that an intervening network has on the measurement of a device under test. De-embedding, on the contrary, is the removal of such effects. When the intervening network is described in terms of S-parameters, as in this paper, the embedding equation corresponds to Equation~\ref{equation_gamma_shifted}.}

Figure~\ref{figure_impedance_balun} shows, for reference, $\Gamma_{A}$ computed using the balun S-parameters from instrument~$1$ measured in the lab and $\Gamma_{Ad}$ simulated with FEKO for the nine soil models (Section~\ref{section_antenna_impedance}). The top row of the figure shows that the balun reduces the reflection magnitude relative to $|\Gamma_{Ad}|$ in Figure~\ref{figure_impedance_port} across most of the frequency range, except for the resonance at $\approx50$--$60$~MHz. As Equation~\ref{equation_gamma_shifted} indicates, when $S^b_{11}$ and $S^b_{22}$ are low, the decrease in the magnitude is mainly driven by $|S^b_{12}S^b_{21}|$, which is $\approx-7$ to $-9$~dB (Figure~\ref{figure_balun1}). The soil differences between our single-layer models produce magnitude variations in the range $\approx-14$ to $-24$~dB at the $\approx50$~MHz dip. For these models, the reflection magnitude increases with both conductivity and permittivity. For our two-layer models, soil changes below the surface lead to changes in reflection magnitude $\lesssim2$~dB.

The bottom row of Figure~\ref{figure_impedance_balun} shows that the reflection phase at the balun output decreases monotonically across $25$--$125$~MHz. The transition at $\approx50$~MHz is smoother than for $\angle\Gamma_{Ad}$ and there are no $360^{\circ}$ jumps. The phase variations across our single-layer (two-layer) models are $\lesssim40^{\circ}$ ($\lesssim10^{\circ}$). The delay of the antenna plus balun across $25$--$125$~MHz is $\approx(630^{\circ}/ 360^{\circ}) \times (100 \;\mathrm{MHz})^{-1} \approx 18\;\mathrm{ns}$. For comparison, the delay of the EDGES-2 low-band antenna including the Roberts balun is $\approx28$~ns \citep{bowman2018}. A lower delay is preferred because it makes the calibrated sky measurements less sensitive to errors in reflection measurements \citep{monsalve2017a}.

\subsection{Balun efficiency}
The balun efficiency is computed as \citep{monsalve2017a} 

\begin{equation}
\eta_{balun} = \frac{|S^b_{12}|^2\left(1-|\Gamma_{Ad}|^2\right)}{|1-S^b_{22}\Gamma_{Ad}|^2\left(1-|\Gamma_{A}|^2\right)}.
\label{equation_balun_efficiency}
\end{equation}

When analysing real measurements, the balun efficiency is computed using the balun S-parameters measured in the lab, $\Gamma_A$ measured in the field (Section~\ref{section_receiver}), and $\Gamma_{Ad}$ obtained by solving Equation~\ref{equation_gamma_shifted} for $\Gamma_{Ad}$. Figure~\ref{figure_balun3} shows, for reference, the balun efficiency computed using $\Gamma_{Ad}$ simulated with FEKO (Section~\ref{section_antenna_impedance}). Panel~(a) shows the efficiency for both MIST instruments and the nominal soil model. For this model, the lowest efficiency is $\approx 5\%$ at $25$~MHz and the highest is $\approx45\%$ at $50$~MHz. The shape of these curves strongly follows the magnitude of the reflection coefficient at the balun output (Figure~\ref{figure_impedance_balun}), with higher efficiency occurring for lower reflection magnitude. Because the antenna reflection coefficient depends on the soil properties, so does the balun efficiency. Panels~(b) and (c) of Figure~\ref{figure_balun3} show the differences in efficiency for the alternative soil models relative to the nominal model with the balun of instrument~$1$. The largest differences occur for the single-layer models and are of up to $\approx8\%$. For the two-layer models, changes in the bottom layer produce efficiency changes of up to $\approx0.8\%$. These differences are significantly larger than the $\lesssim0.01\%$ ratio between global $21$~cm signal and astrophysical foreground, and have a spectral structure that could bias the $21$~cm signal extraction if not accounted for. It is thus necessary to compute the balun efficiency using high-accuracy {\it in situ} measurements of the antenna reflection coefficient.

\section{Receiver}
\label{section_receiver}

\begin{figure*}
\centering
\includegraphics[width=\linewidth]{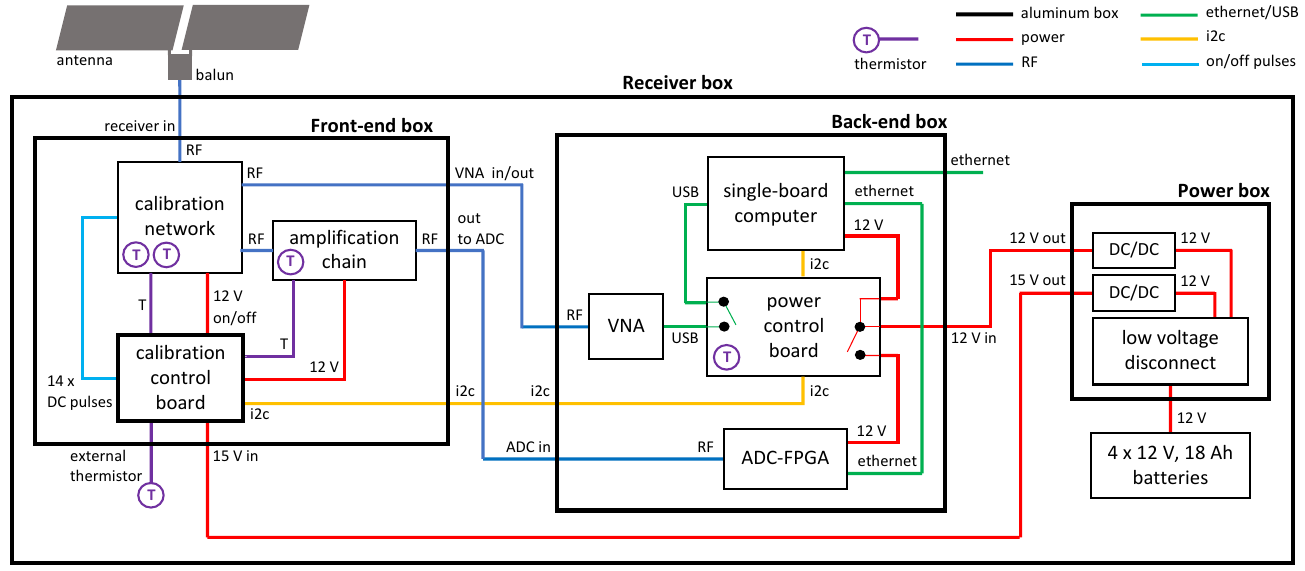}
\caption{Block diagram of the MIST receiver.}
\label{figure_block_diagram}
\end{figure*}

\begin{figure*}
\centering
\includegraphics[width=\linewidth]{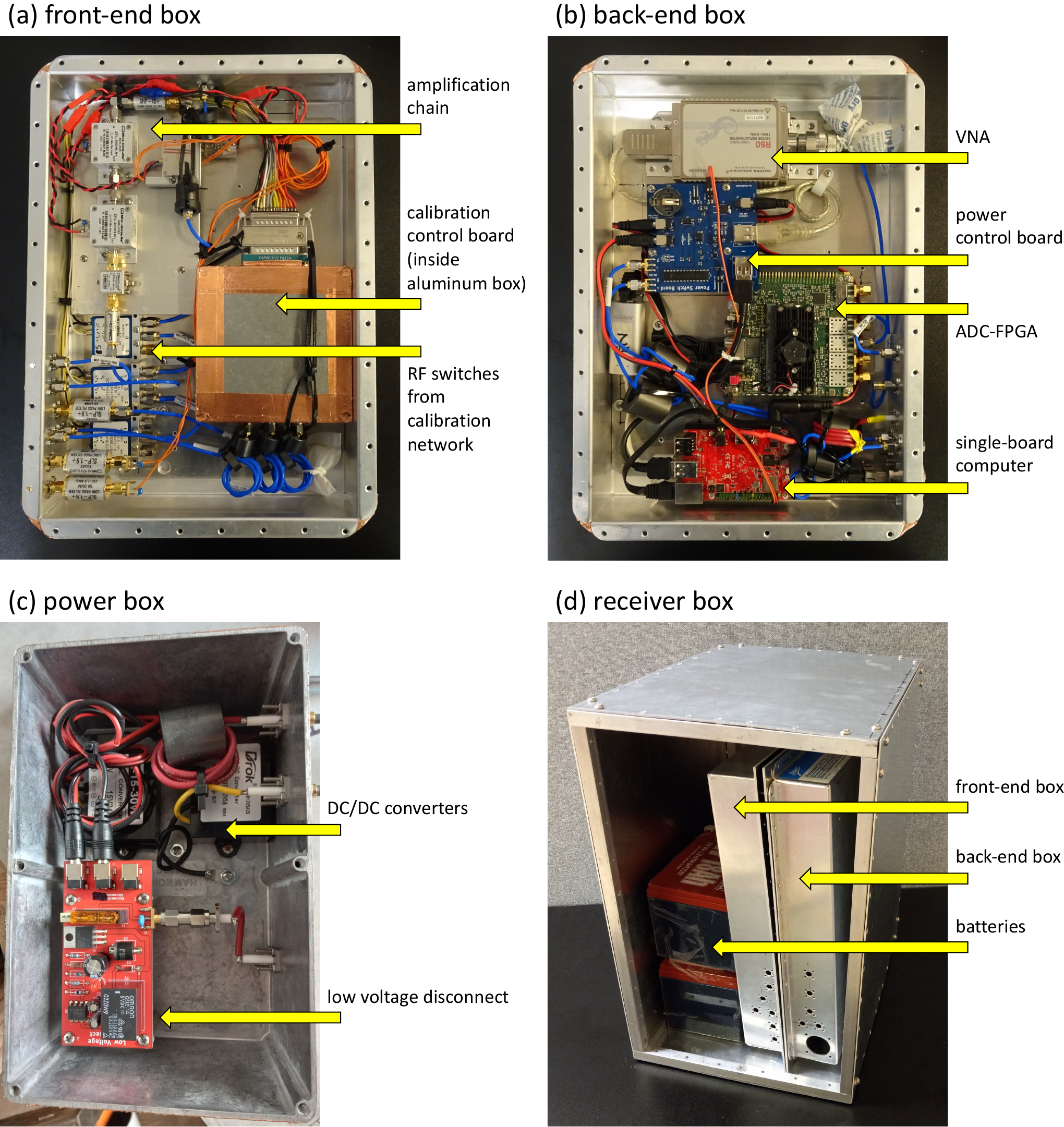}
\caption{One of the MIST receivers: (a) front-end box; (b) back-end box; (c) power box; and (d) receiver box. The picture of the receiver box was taken during construction and assembly, and not all the features are shown. Three specific things not shown are: (1) the SMA input connector at the top of the box; (2) the plastic sleeves that protect the batteries; and (3) the power box, which sits above the batteries.}
\label{figure_instrument_box}
\end{figure*}

In this section, we describe the hardware and operation of the MIST receiver.

\subsection{Overview}
Figure~\ref{figure_block_diagram} shows a block diagram of the receiver. Inside the receiver box, the electronics are housed in three boxes: the front-end box, the back-end box, and the power box. Figure~\ref{figure_instrument_box} shows pictures of these boxes.

After leaving the balun, the antenna signal enters the front-end box. Here, the first subsystem encountered is the calibration network. This network contains calibration standards necessary for the implementation of our receiver calibration formalism (Section~\ref{section_formalism_receiver}). The calibration network has two outputs; one output to measure the PSDs of the antenna and the internal PSD calibration devices, and the other output to measure the reflection coefficients of the antenna and the internal PSD calibration devices, as well as the reflection coefficient looking into the receiver input. From the PSD output,\footnote{Although when leaving the calibration network the signal is an analogue RF signal, we call this output the `PSD output' because it provides the signal from which we compute the PSD in the back-end box.} the signal goes through an amplification chain and is then routed to the back-end box. From the reflection output, the signal is routed directly to the back-end box. A calibration control board in the front-end box is used to provide the control signals required by the calibration network.

In the back-end box, the analogue signal from the PSD output is digitized by an ADC and transformed to the frequency domain by a field-programmable gate array (FPGA). A single ADC-FPGA board performs both functions. The signal from the reflection output is measured by a vector network analyser (VNA). A power control board is used to independently turn on and off the ADC-FPGA and VNA. A single-board computer (SBC) coordinates the PSD and reflection measurements, and also stores the data. To coordinate the measurements, the SBC controls the power control board in the back-end box and the calibration control board in the front-end box. An ethernet connection is used to communicate with the SBC from outside the back-end box. 

The instrument is powered by four $12$~V, $18$~Ah batteries. Inside the power box, two DC/DC converters stabilize the fluctuating battery voltage to $15$~V and $12$~V. These voltages are used to power the front-end and back-end boxes, respectively. The power box also contains a low voltage disconnect (LVD) circuit to prevent the discharging of the batteries below safe levels.

\begin{figure}
\centering
\includegraphics[width=\linewidth]{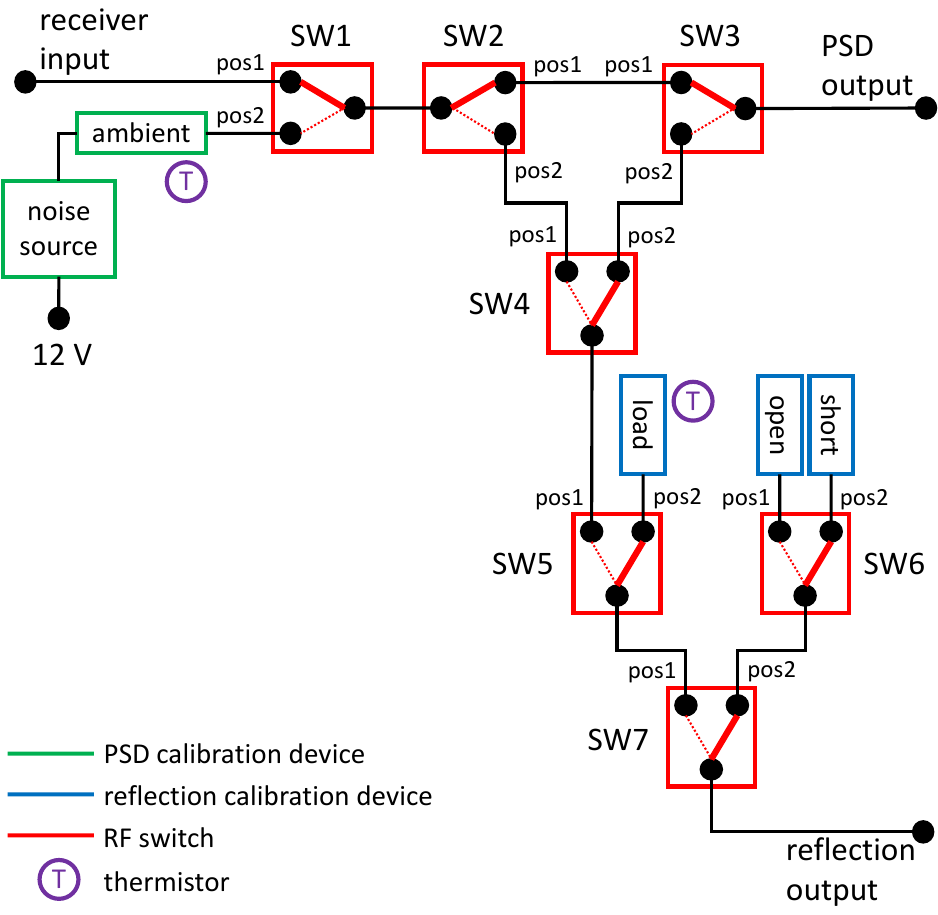}
\caption{Diagram of the calibration network. The network is composed of calibration standards for PSD and reflection measurements, latching RF switches, hand-formable coaxial cables for RF connections within and from the network, and two thermistors.}
\label{figure_calibration_network}
\end{figure}

\subsection{Front-end box}
\subsubsection{Calibration network}
\label{section_calibration_network}

Figure~\ref{figure_calibration_network} shows the components of the calibration network. The network contains an ambient load and an active noise source for the calibration of PSD measurements (Section~\ref{section_formalism_receiver}). The ambient load is implemented as two SMA attenuators ($20$ and $6$~dB) connected in series for a total attenuation of $26$~dB. These attenuators provide a noise temperature equivalent to their physical temperature (typically $\approx300$~K). The noise source is built around a noise diode with an excess noise ratio of $\approx35$~dB. When combined, the ambient load plus noise source provide a nominal noise temperature of $\approx$2,500~K, which is higher than the expected antenna temperature from sky measurements affected by losses (Equation~\ref{equation_antenna_temperature1}). The noise temperature of the ambient load plus noise source was tuned to optimize the dynamic range of the ADC. The noise source is powered with $12$~V and only during the measurement of its PSD. The calibration network also contains open, short, and $50$~$\Omega$ load (OSL) standards for the calibration of reflection coefficient measurements. For the open and short we use commercially available SMA caps. The load is implemented using a $50$~$\Omega$ resistor that is temperature-stable to better than $10$~ppm~$^{\circ}$C$^{-1}$. We track the physical temperatures of the $20$+$6$~dB attenuators and the $50$~$\Omega$ resistor using $15$~k$\Omega$ thermistors.

The calibration network contains seven radio frequency (RF) switches (SW1 through SW7) to route the signals in the required directions. The RF switches are electromechanical, single-pole-double-throw (SPDT), latching switches with SMA connectors. Switch SW1 selects between the receiver input (to which we connect the antenna plus balun or an external calibration device), and the internal ambient load and noise source. SW2 routes the output of SW1 either toward the PSD output or the reflection output. SW3 is used to switch the input of the amplification chain between two modes: PSD measurements through SW1 and SW2, and the measurement of the reflection coefficient looking into the amplification chain through SW4. Switch SW4 is also used to measure the reflection coefficient of the antenna, internal ambient, and ambient plus noise source, through SW1 and SW2. Switches SW5, SW6, and SW7 are used to measure the reflection coefficient of the OSL standards. The RF connections between switches are done using hand-formable coaxial cables from the Mini-Circuits 086-XSM+ series, most of which are $4$-inch long. The RF connections from the calibration network toward the receiver input, amplification stage, and VNA, are done using the same type of cable.

The design of the MIST calibration network was influenced by EDGES-2, which was the first global 21~cm experiment that incorporated hardware to autonomously measure the reflection coefficient of the antenna at the receiver input \citep{monsalve2017a,bowman2018}. As described in the previous paragraph, in addition to the antenna, MIST autonomously measures the reflection coefficient of the internal ambient load and ambient plus noise source, used for PSD calibration, as well as the reflection coefficient looking into the receiver input. These additional {\it in situ} measurements are conducted to increase the calibration accuracy and track the instrument's performance in the field.

The current EDGES design, EDGES-3, incorporates hardware to conduct more autonomous calibration measurements than EDGES-2 and MIST \citep{rogers2019}. This hardware, located inside the antenna, enables EDGES-3 to autonomously perform the full set of PSD and reflection coefficient measurements required to determine the absolute receiver calibration parameters ($C_1$, $C_2$, $T_U$, $T_C$, and $T_S$, Section~\ref{section_formalism_receiver}). REACH also incorporates hardware to autonomously conduct a full receiver calibration in the field \citep{deleraacedo2022}. In MIST, we carry out the absolute receiver calibration measurements by manually connecting to the receiver input the calibration devices on the outside of the receiver box (Sections~\ref{section_formalism_receiver} and \ref{section_calibration}). This approach was adopted to keep the MIST receiver small and considering that the receiver parameters can be determined in the lab by controlling the physical temperature of the electronics to match the operation temperature in the field.

In their calibration networks, EDGES-2, EDGES-3, and REACH use different types of RF switches, including switches with more than two positions. As described before and shown in Figure~\ref{figure_calibration_network}, MIST only uses SPDT switches, which are interconnected with hand-formable coaxial cables. SPDT switches are the most common type of electromechanical RF switch. Using only one and, in particular, the most common type of electromechanical switch, makes it easier to replace switches in the field in case of malfunction.

\begin{figure}
\centering
\includegraphics[width=\linewidth]{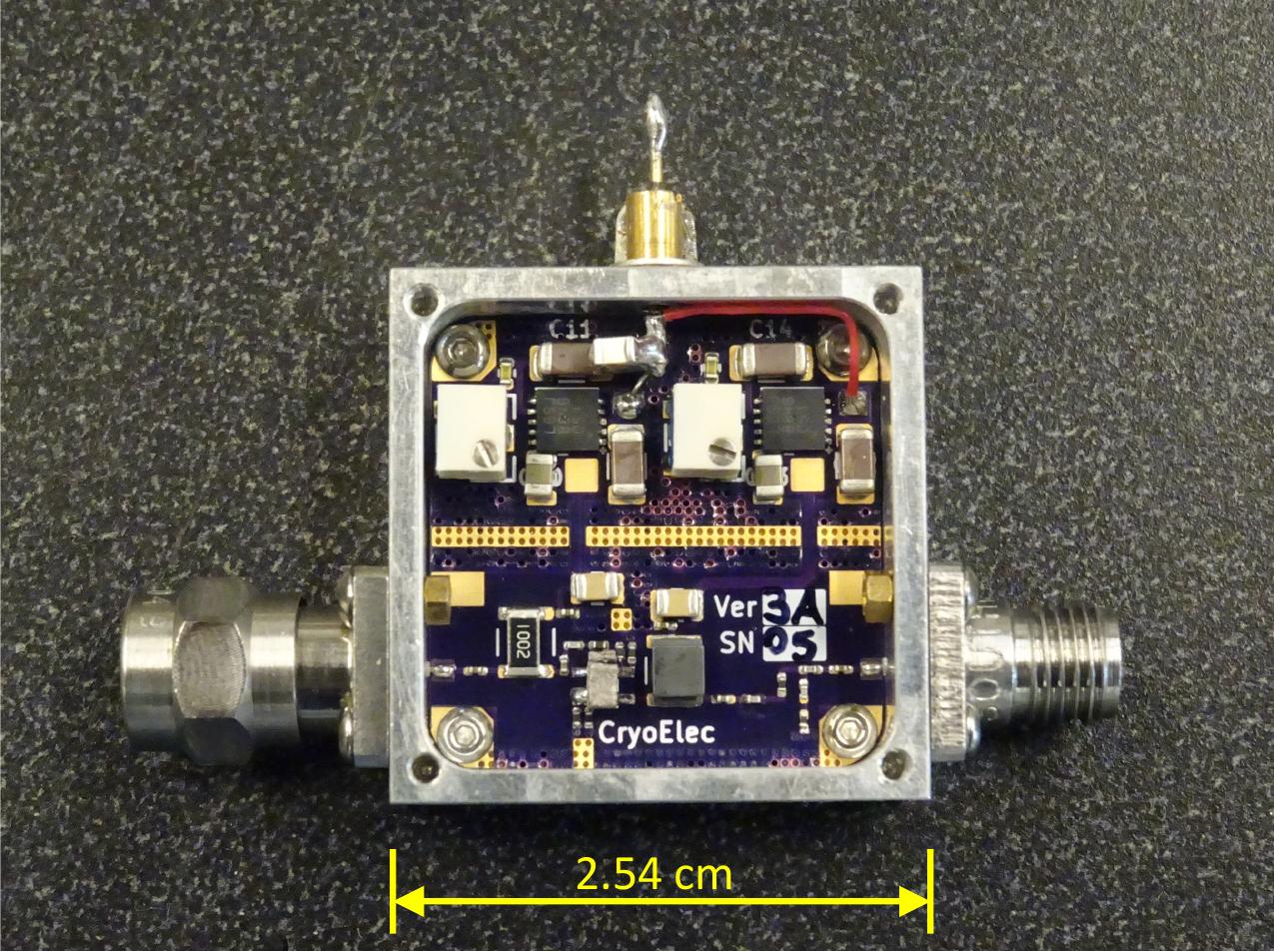}
\caption{Custom LNA used by MIST.}
\label{figure_LNA}
\end{figure}

\subsubsection{Amplification chain}
In PSD measurement mode, after leaving the calibration network, the signal from the receiver input (or internal ambient and noise source) reaches the amplification chain. The first elements in the chain are a $3$~dB attenuator and an LNA. The attenuator is used to improve the impedance match between the LNA and the antenna. The increased noise temperature due to this attenuator is small compared to the temperature from the astrophysical foregrounds which, aside from RFI, is the largest contributor to the system temperature when the receiver input is in the antenna position. Figure~\ref{figure_LNA} shows a picture of the LNA. The LNA is a custom-made, silicon-germanium, heterojunction bipolar transistor design optimized for low reflections at the input and output ($|S_{11}|$ and $|S_{22}|<-35$~dB), as well as for low reverse transmission ($|S_{12}|<-40$~dB). The gain of the LNA is $\approx28$~dB and its noise temperature is $\approx150$~K.  We measure the physical temperature of the LNA using a thermistor. 

A $10$~dB attenuator is connected to the output of the LNA to improve the match between the LNA and the next components, as well as to regulate the total receiver gain. After the $10$~dB attenuator, a $48$~MHz high-pass filter is used to reduce incoming power from strong shortwave RFI. This filter is followed by two $\approx20$~dB Mini-Circuits ZFL-HLN+ amplifiers, another $48$~MHz high-pass filter, and a $120$~MHz low-pass filter used to suppress RFI above the band, such as from ORBCOMM satellites, and contamination from aliased signals. We have not encountered the need to inject noise to the signal chain below the observation band (i.e. $<25$~MHz) for conditioning purposes, as performed by EDGES \citep{rogers2012,monsalve2017a,bowman2018}. The total nominal gain of the amplification chain is $\approx55$~dB. The power consumption of the amplification chain is $2.5$~W.

\subsubsection{Calibration control board}
The calibration control board is used to generate the pulses that control the latching RF switches in the calibration network. These pulses have a voltage of $12$~V, a duration of one second, and are produced using MOSFET relays. The calibration control board also regulates and filters the $15$~V coming from the power box to produce the $12$~V used for the pulses. These $12$~V are also used to power the amplification chain and the noise diode. The calibration control board includes low-speed ADCs, which connect to the thermistors that measure the physical temperatures of the LNA, the internal calibration loads, and the external calibration devices.

\subsection{Back-end box}
\subsubsection{ADC-FPGA}

The ADC-FPGA used by MIST is a Koheron ALPHA250,\footnote{\url{https://www.koheron.com/fpga/alpha250-signal-acquisition-generation}} which is powered with $12$~V and has a consumption of $10$~W. The board has two RF ADCs and two RF digital-to-analogue converters. The two ADCs correspond to two channels of the Linear Technologies LTC2157-14 chip. MIST uses only one ADC. The ADCs have a sampling rate of $250\times10^6$~samples per second and an amplitude resolution of $14$~bits. The signal-to-noise plus distortion ratio of the ADCs is $69$~dBFS and the effective number of bits is $11.2$. The integral and differential linearity errors of the ADCs are $\pm0.85$~LSB and $\pm0.25$~LSB, respectively. The spurious-free dynamic range of the ADCs for input signals in the range $0$--$125$~MHz is $85$~dBFS.

The FPGA side of the ADC-FPGA board is built around a Zynq~7020 system-on-a-chip, which integrates an ARM Cortex-A9 processor and a Xilinx 7-series FPGA. The FPGA converts the digital signal to the frequency domain by applying an FFT algorithm to non-overlapping time windows multiplied by a Blackman-Harris window function. Each FFT computation uses 8,192 time samples and produces a spectrum in the range $0$--$125$~MHz with a resolution of $30.518$~kHz. Simulations indicate that the time efficiency introduced by the Blackman-Harris window function relative to a rectangular window function is $36\%$.

\subsubsection{Vector network analyser}
The VNA used by MIST is a Copper Mountain Technologies R60,\footnote{\url{https://coppermountaintech.com/vna/r60-1-port/}} which has one port and a frequency range of $1$~MHz--$6$~GHz. This VNA is powered through its USB connection and has a consumption of $3$~W. For reference, the manufacturer's nominal accuracy specifications over the full $1$~MHz--$6$~GHz range go from $\pm0.2$~dB / $\pm2^{\circ}$ for a reflection magnitude of $0$~dB to $\pm3.0$~dB / $\pm18^{\circ}$ for a magnitude of $-35$~dB. In MIST, the reflection measurements with the R60 VNA are done over the range $1$--$125$~MHz, with a resolution of $250$~kHz, and an intermediate frequency bandwidth of $100$~Hz. These settings produce low measurement noise and a sweep time of $6$~s. The measurements are calibrated using the techniques described in Section~\ref{section_reflection_coefficient} to achieve an accuracy of $\mathcal{O}(10^{-4})$.

\subsubsection{Power control board}
The power control board is used to turn on and off the ADC-FPGA and VNA. This switching is done using MOSFET relays. A thermistor and low-speed ADC are included in this board to monitor the air temperature in the back-end box.

\subsubsection{Single-board computer}
The SBC used by MIST is a Radxa ROCK~Pi~X,\footnote{\url{https://wiki.radxa.com/RockpiX}} which is designed around an Atom x5-Z8350 processor. The SBC is powered with $12$~V and has a consumption of $4.5$~W. In the receiver, the SBC communicates with the ADC-FPGA using ethernet; with the VNA using USB; and with the power control board and calibration control board using I$^2$C. The communication with an external laptop, to start and stop observations as well as to retrieve data, is done through a second ethernet connection.

\subsection{Power}
The power consumption of the receiver during PSD (VNA) measurements is $17$~W ($10$~W). This low consumption enables the instrument to operate for significant periods of time with small batteries. We use four $12$~V, $18$~Ah batteries connected in parallel, which provide enough capacity to conduct continuous autonomous operations for $48$~h (without fully discharging the batteries for their protection). The size of each battery is $10$~cm~$\times$~$10$~cm~$\times$~$15$~cm. The weight of each battery is $2$~kg. The batteries are sufficiently small that they can be housed in the receiver box directly underneath the antenna (Figure~\ref{figure_instrument_box}), thus increasing instrument portability. The internally housed batteries also eliminate potential systematic effects associated with long cables connecting the instrument to an external power source.

\subsection{Receiver box and suppression of self-RFI}
\label{section_self_RFI}

Because the receiver box is in close proximity to the antenna, multiple layers of self-RFI suppression are required. The receiver box consists of walls made of aluminum sheet attached with stainless steel screws to a frame made of aluminum bars. The screws are placed every $3$~cm along the perimeter of the walls. Inside the box, copper tape is used along the edges to maximize the electrical contact between the walls and the frame. The receiver input connector is located at the top of the box and consists of a female-female flange mount SMA adapter. On the outside of the box, this adapter connects directly to the balun. A ferrite bead is used on this SMA connection to suppress common-mode currents. The box does not have any other pass-through connection. To communicate with the instrument, as well as to replace the batteries, the front wall of the box has to be unscrewed and removed.

The front-end, back-end, and power boxes, including their lids, are made of aluminum. The lids of the front- and back-end boxes are secured with stainless steel screws, which are also placed every $3$~cm along the perimeter. Braided metal gasket and copper tape are also used along the perimeter of the lids to ensure full electrical contact with the boxes. Inside the front-end box, the calibration control board is housed in its own aluminum box for suppression of potential RFI from the I$^2$C bus and peripherals. An RJ45 feedthrough connector on the back-end box enables the ethernet communication between the SBC and an external laptop. This connector has a metal cap on the outside, which is bolted on during measurements. 

Ferrite beads and capacitive filters are used on select connections inside the front-end, back-end, and power boxes. Ferrite beads are also used on the connections between the three boxes, as well as between the power box and the batteries. All the contents of the receiver box are wrapped with aluminum foil for additional RFI suppression.

\begin{table}
\caption{Structure of the MIST measurement blocks. The duration of the measurements reported here includes buffer time needed to switch between positions. The total duration for the $160$ PSD cycles (cycle~\#~$3$ through $162$) is $160\times41$~s~$=109$~min. The total duration of the measurement blocks is $111$~min.}             % title of Table
\label{table_measurement_block}      % is used to refer this table in the text
\centering                          % used for centering table
\begin{tabular}{c l l c}        % centered columns (4 columns)
\hline % inserts double horizontal lines
\\
Cycle \# &Measurement type &  Device & Duration \\ % table heading 
\hline  
$1$ & Reflection $0$~dBm  &  & 56~s   \\
\hline                        % inserts single horizontal line
&&Thermistors\tablefootnote{For the thermistors we measure their voltage, instead of reflection coefficient or PSD.\\}             & 4~s \\
&&O                       & 8~s      \\
&&S                       & 8~s      \\
&&L                       & 8~s      \\
&&Antenna                       & 8~s      \\
&&Ambient                       & 8~s      \\
&&amb+ns                       & 12~s      \\
\\
\hline                        % inserts single horizontal line
$2$ & Reflection $-40$~dBm  &  & 51~s      \\
\hline
&&Thermistors             & 4~s     \\
&&O                       & 8~s     \\
&&S                       & 8~s     \\
&&L                       & 8~s     \\
&&Receiver input          & 23~s\tablefootnote{This duration includes $15$~s in which, after measuring the reflection looking into the receiver input, the VNA is turned off, the ADC-FPGA is turned on, and the instrument waits until the ADC-FPGA is ready to take data.}     \\
\\
\hline                        % inserts single horizontal line
$3$--$162$ & PSD         &           & 41~s     \\
\hline
&&Thermistors             & 4~s      \\
&&Antenna                 & 12~s     \\
&&Ambient                 & 12~s     \\
&&amb+ns                  & 13~s     \\
\\
\hline                                   %inserts single line
\end{tabular}
\end{table}

% \tablefoot{\\
%\tablefoottext{a}{For the thermistors we measure their voltage, instead of reflection coefficient or PSD.\\}
%\tablefoottext{b}{This duration includes $15$~s in which, after measuring the reflection looking into the receiver input, the VNA is turned off, the ADC-FPGA is turned on, and the instrument waits until the ADC-FPGA is ready to take data.}}\\

\subsection{Measurement sequence}
\label{section_measurements}
The MIST instrument takes measurements of the sky and internal calibration standards in blocks of $111$~min. The structure of these blocks is shown in Table~\ref{table_measurement_block}. Each block is organized into $162$ cycles. In cycle~\#~$1$, the instrument measures the reflection coefficient of six devices with a VNA power of $0$~dBm for a total of $56$~s. Here, the main measurement is of the antenna or external device connected to the receiver input. We also measure the internal ambient load and ambient plus noise source to verify the correct functioning of the instrument. The internal OSL standards are measured to calibrate the other three measurements. In cycle~\#~$2$, which lasts $51$~s, the instrument measures the reflection coefficient of four devices at $-40$~dBm. Here, the main measurement is the reflection looking into the receiver input, which has to be done at a lower power to avoid saturating the LNA. The internal OSL standards are also measured at this lower power for consistency of the calibration. Starting with cycle~\#~$3$, the instrument carries out $160$ cycles of PSD measurements for a total of $109$~min. Three measurements are done in each PSD cycle: (1) the antenna, (2) internal ambient load, and (3) ambient plus noise source. Each of these measurements corresponds to a $10$-s integration. At the beginning of each reflection and PSD cycle, the instrument measures the physical temperatures of all the thermistors. After a $111$-min block finishes, a new one automatically starts. This process continues until a stop command is sent to the instrument from the external laptop, or the battery voltage decreases below the LVD threshold. 

Data are saved as binary files on the flash memory of the SBC. We save one file per measurement block. This file is continuously updated as new measurements are captured. The file size for a full block is $16.2$~MB, yielding a daily data rate of $210$~MB.

\begin{figure}
\centering
\includegraphics[width=\linewidth]{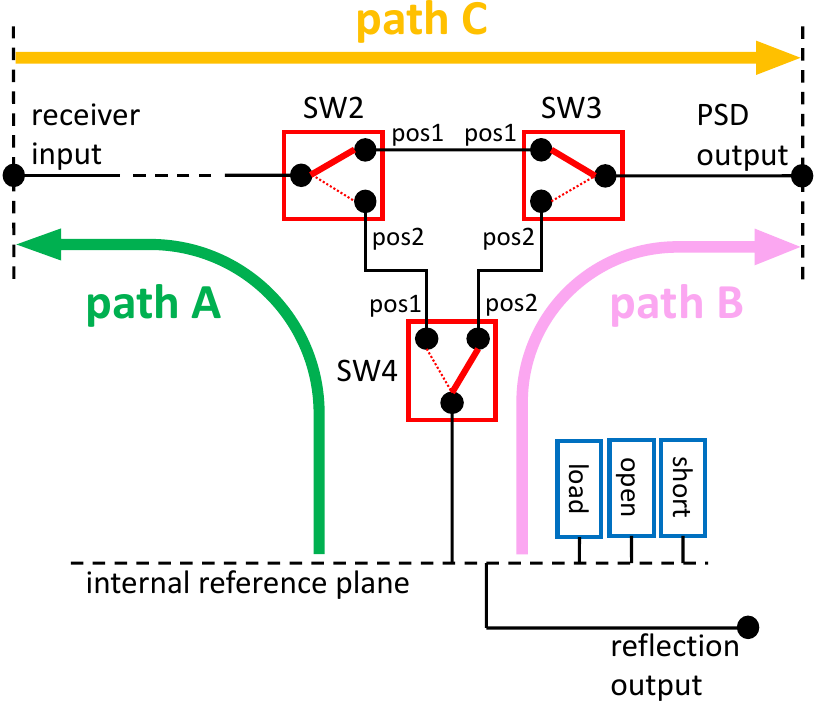}
\caption{Diagram of the calibration network highlighting the paths that shift the internal reflection reference plane to the receiver input.}
\label{figure_calibration_network2}
\end{figure}

\begin{figure}
\centering
\includegraphics[width=\linewidth]{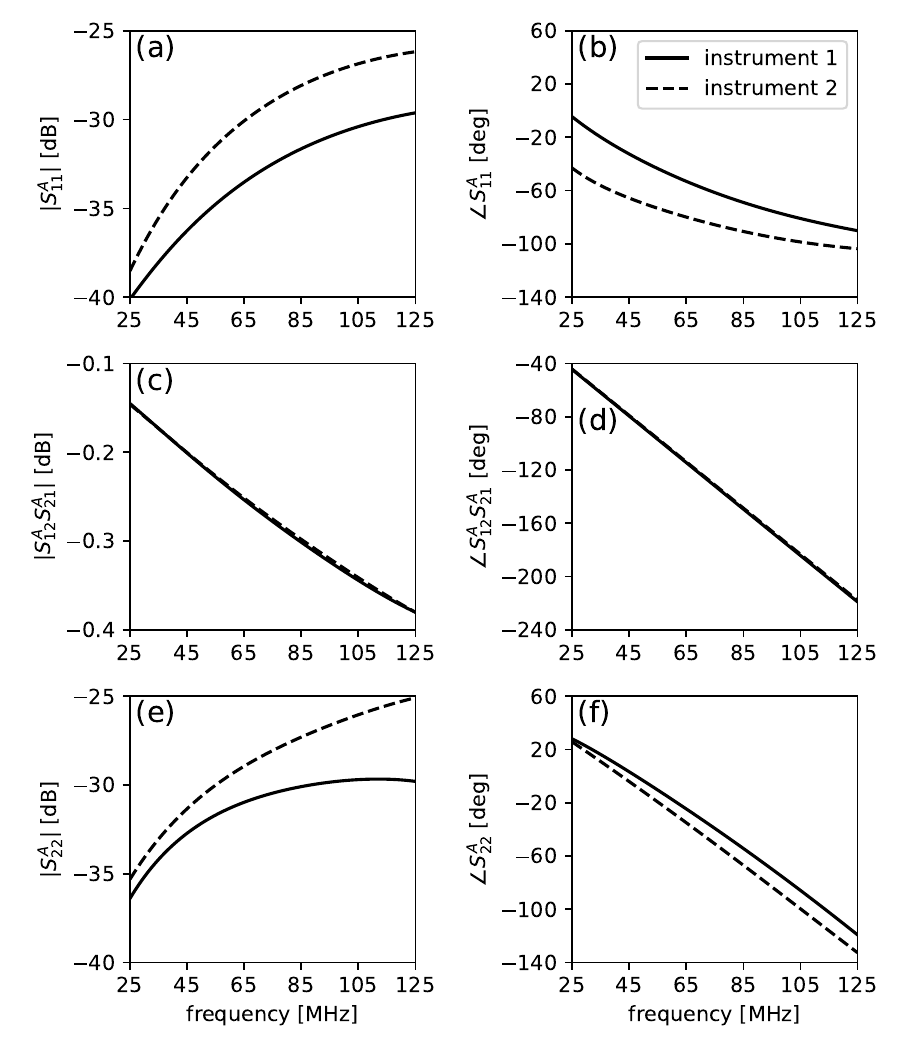}
\caption{S-parameters of path~A in the calibration network measured in the lab. Port~$1$ corresponds to the internal reflection reference plane and port~$2$ to the receiver input.}
\label{figure_sparameters_pathA}
\end{figure}

\subsection{Reflection coefficient calibration}
\label{section_reflection_coefficient}
The reflection measurements described in Section~\ref{section_measurements} are done using the calibration network and VNA incorporated into the receiver. Our receiver calibration formalism (Section~\ref{section_formalism_receiver}) requires these reflection measurements themselves to be calibrated at the receiver input. Furthermore, and as discussed in \citet{monsalve2017a,monsalve2017b}, the detection of the global $21$~cm signal requires an accuracy in reflection coefficient measurements of $\mathcal{O}(10^{-4})$. To calibrate the reflection measurements at the receiver input and with the required accuracy, MIST conducts the two-step process used by EDGES \citep{monsalve2017a,bowman2018}. First, a relative calibration is done using the measurements of the internal OSL standards. This step de-embeds a large fraction of the unwanted S-parameters introduced by the calibration network, the cable to the VNA, and the VNA itself. Calibrating using the internal OSL standards defines the reference plane within the calibration network. Second, the reference plane is shifted from the calibration network to the receiver input by applying S-parameter corrections determined in the lab. These corrections account for the short paths between the internal reference plane and the receiver input.

To shift the reference plane to the receiver input, three paths have to be accounted for and characterized. The paths, labelled A, B, and C, are shown in Figure~\ref{figure_calibration_network2}. Once the S-parameters of the paths are available, the measurements of external devices (such as the antenna) are calibrated at the receiver input by: (1) doing a relative calibration using the internal OSL standards, and (2) de-embedding the S-parameters of path~A. Similarly, the measurement of reflection coefficient looking into the receiver input is calibrated by: (1) doing a relative calibration using the internal OSL standards; (2) de-embedding the S-parameters of path~B, which shifts the calibration plane to the PSD output; and (3) embedding the S-parameters of path~C, which shifts the calibration plane to the receiver input.

Figure~\ref{figure_sparameters_pathA} shows the S-parameters for path A. The parameters for paths B and C are similar and not shown for brevity. These parameters were measured in the lab and calibrated in the same way as the S-parameters of the balun (Section~\ref{section_balun_description}).

\begin{figure}
\centering
\includegraphics[width=\linewidth]{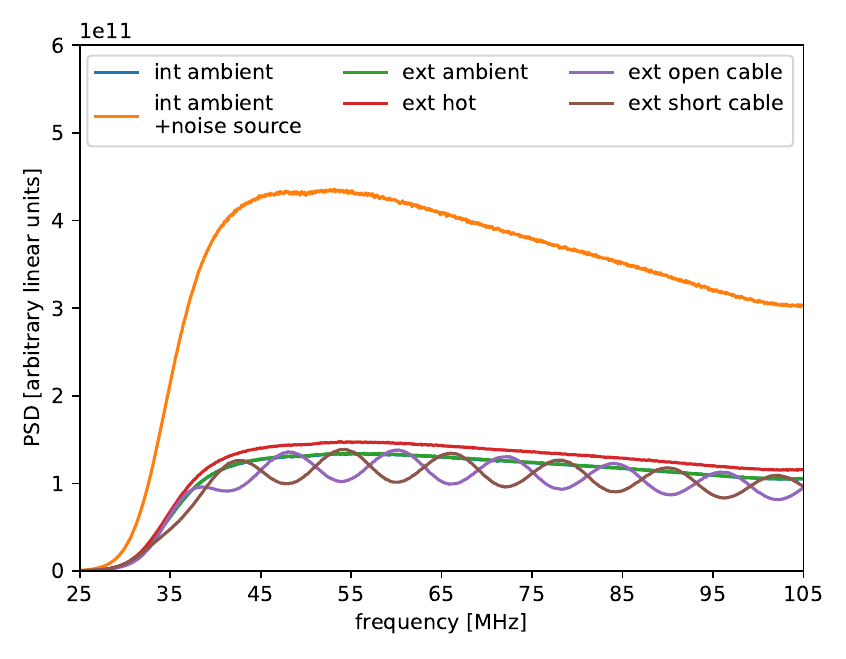}
\caption{Sample PSDs from the internal and external calibrators measured in the lab with instrument~1. PSDs from instrument~2 are similar. Each PSD corresponds to a $10$-s integration.}
\label{figure_psd}
\end{figure}

\begin{figure}
\centering
\includegraphics[width=\linewidth]{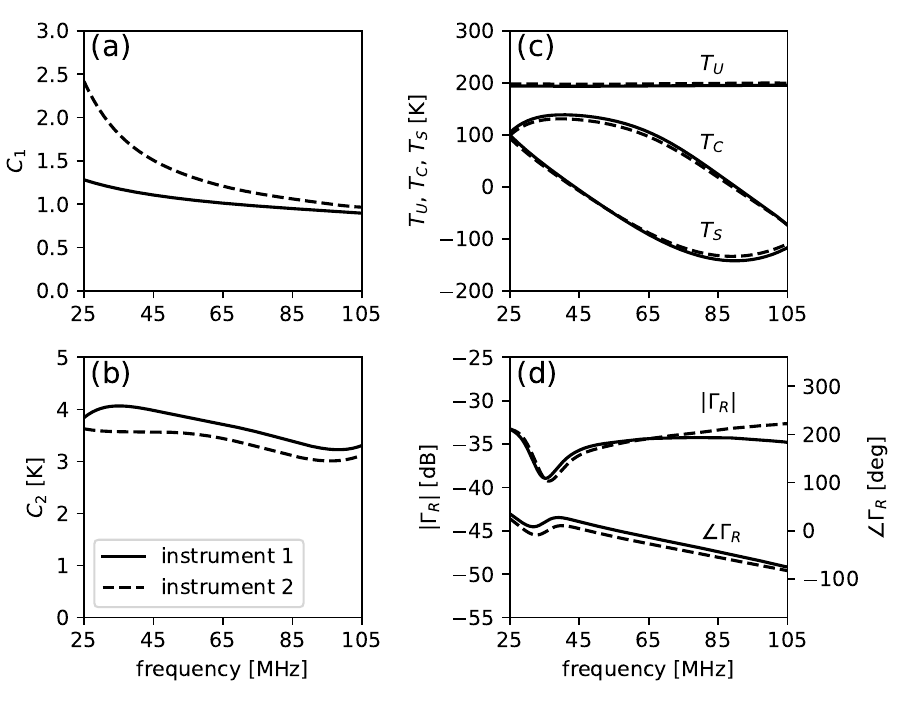}
\caption{(a)-(c) Absolute receiver calibration parameters determined from measurements of external calibration devices conducted in the lab. (d) Reflection coefficient looking into the receiver input measured in the lab.}
\label{figure_calibration_parameters}
\end{figure}

\subsection{Receiver calibration}
\label{section_calibration}

We determine the five absolute receiver calibration parameters, $C_1$, $C_2$, $T_U$, $T_C$, and $T_S$ (Section~\ref{section_formalism_receiver}), from lab measurements of four external calibration devices: (1) an ambient load, (2) a hot load ($50$~$\Omega$ resistor heated up to $\approx400$~K), (3) a $10$~m open-ended low-loss cable, and (4) the same cable after being short-circuited at its far end. For each of these devices, we measure the reflection coefficient, PSD, and physical temperature following the same procedure as for the antenna in the field (Section~\ref{section_measurements}). Figure~\ref{figure_psd} shows sample PSDs from the internal and external calibrators measured during this process. With these measurements at hand, the five receiver parameters are computed using the iterative method of \citet{monsalve2017a}. In the computation, we use $300$~K and 2,300~K as the assumptions for the noise temperature of the internal ambient load ($T^a_L$) and ambient load plus noise source ($T^a_{L+NS}$), respectively. Figure~\ref{figure_calibration_parameters} shows the five parameters, as well as the reflection coefficient looking into the receiver input, for the two receivers.

\begin{figure*}
\centering
\includegraphics[width=\linewidth]{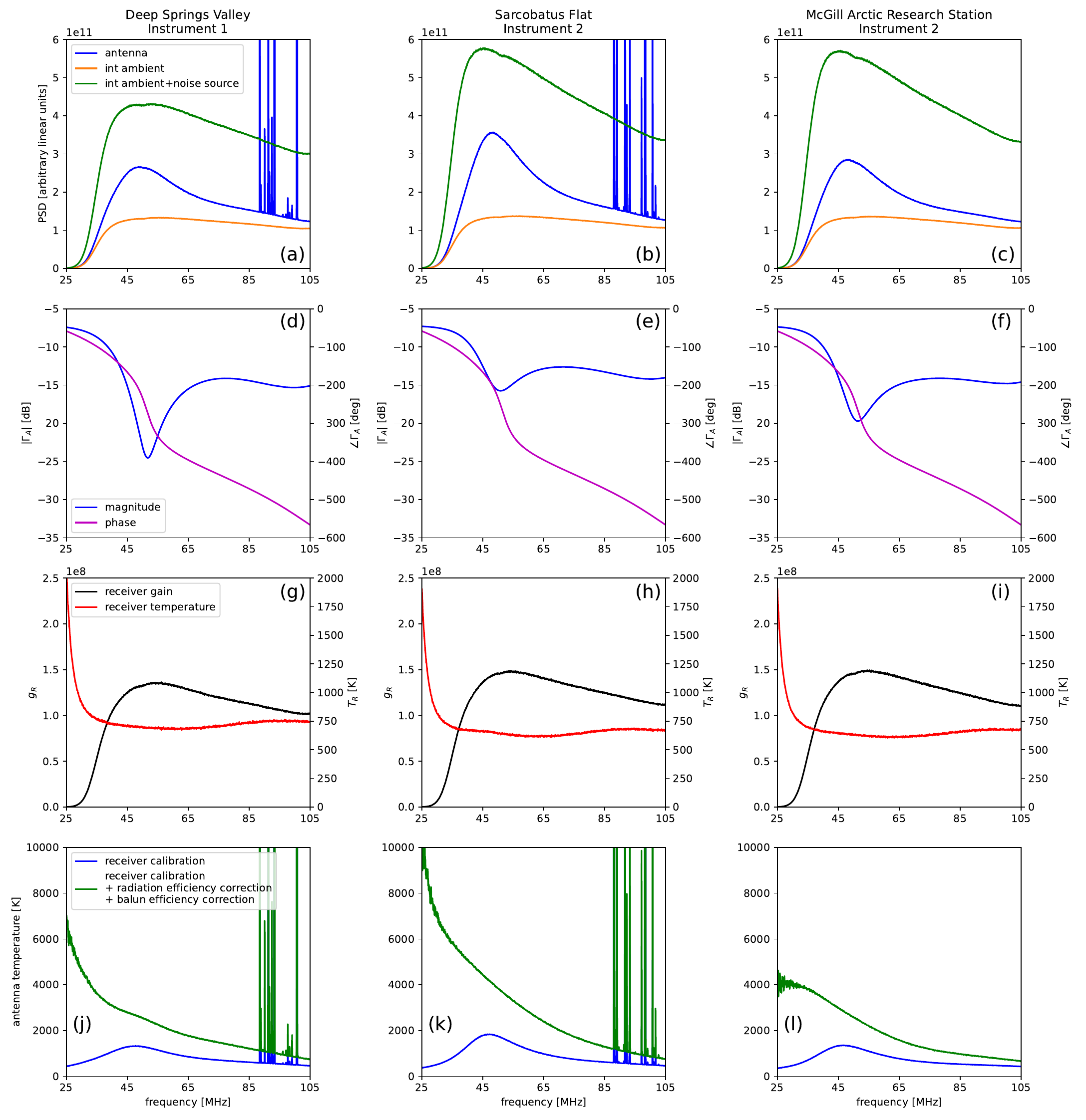}
\caption{Sample measurements from MIST taken at the three observation sites. These measurements preliminarily show that MIST is performing as expected. The columns correspond to data from: (\textit{left}) Deep Springs Valley taken with instrument~$1$; (\textit{middle}) the Sarcobatus Flat taken with instrument~$2$; and (\textit{right}) MARS taken with instrument~$2$. (\textit{Top row}) PSDs from the antenna and internal calibrators corresponding to $10$-s integrations taken at an LST=$15$~h. RFI has not been excised from the antenna measurements, which highlights the excellent radio-quiet conditions at MARS. (\textit{Second row}) Antenna reflection coefficient at the balun output. (\textit{Third row}) Receiver gain and temperature. (\textit{Bottom row}) $41$-second antenna temperature spectra after (1) receiver calibration, and (2) receiver calibration plus radiation and balun efficiency corrections.}
\label{figure_field}
\end{figure*}

\begin{figure}
\centering
\includegraphics[width=0.47\textwidth]{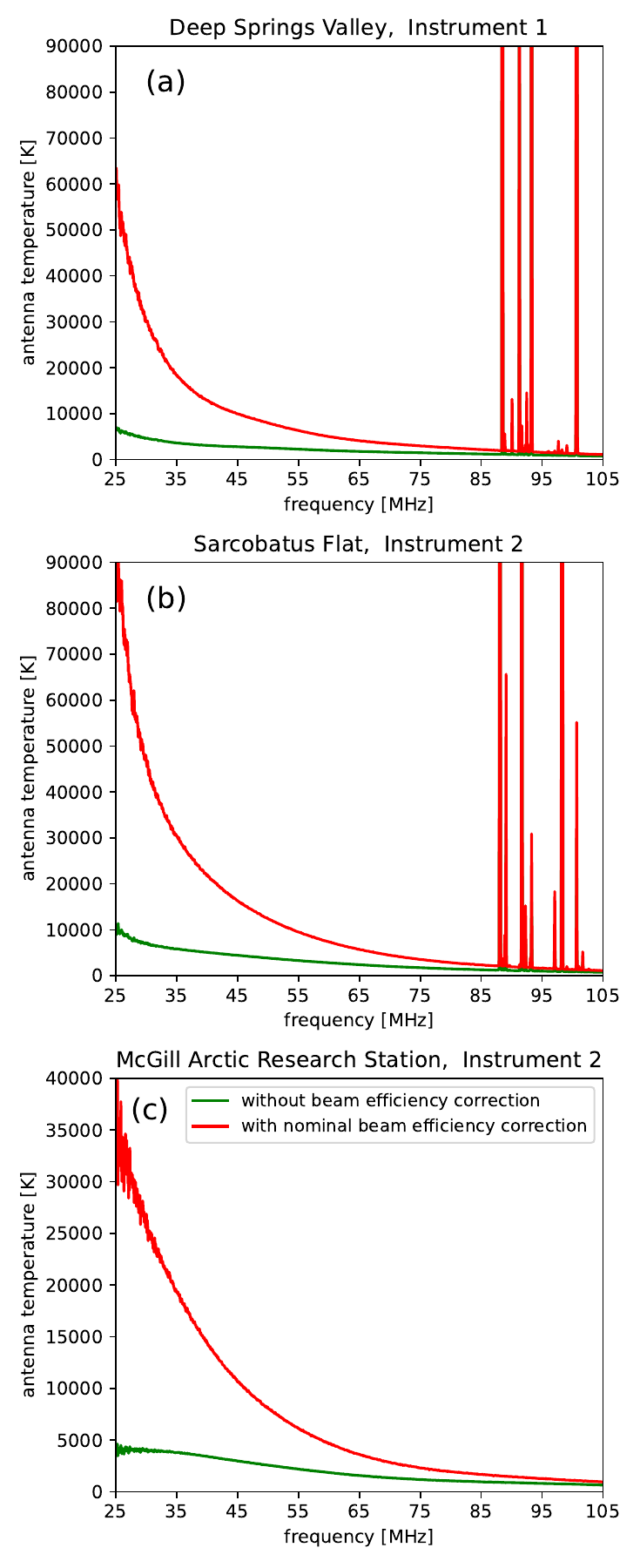}
\caption{Sample $41$-s spectra from Deep Springs Valley, the Sarcobatus Flat, and MARS, before and after beam efficiency correction. RFI has not been excised from these measurements, which highlights the excellent radio-quiet conditions at MARS. In panel~(c), the upper limit of the $y$-axis was reduced to $40000$~K to show the data from MARS in more detail. The beam efficiency used here corresponds to the nominal soil model (Table~\ref{table_soil_parameters}). This beam efficiency is used only as an example. In the future, we will refine the soil model based on {\it in situ} measurements of the antenna reflection coefficient. The spectra without correction (green lines) are reproduced from Figure~\ref{figure_field}, panels~(j), (k), and (l).}
\label{figure_field2}
\end{figure}

\begin{figure}
\centering
\includegraphics[width=\linewidth]{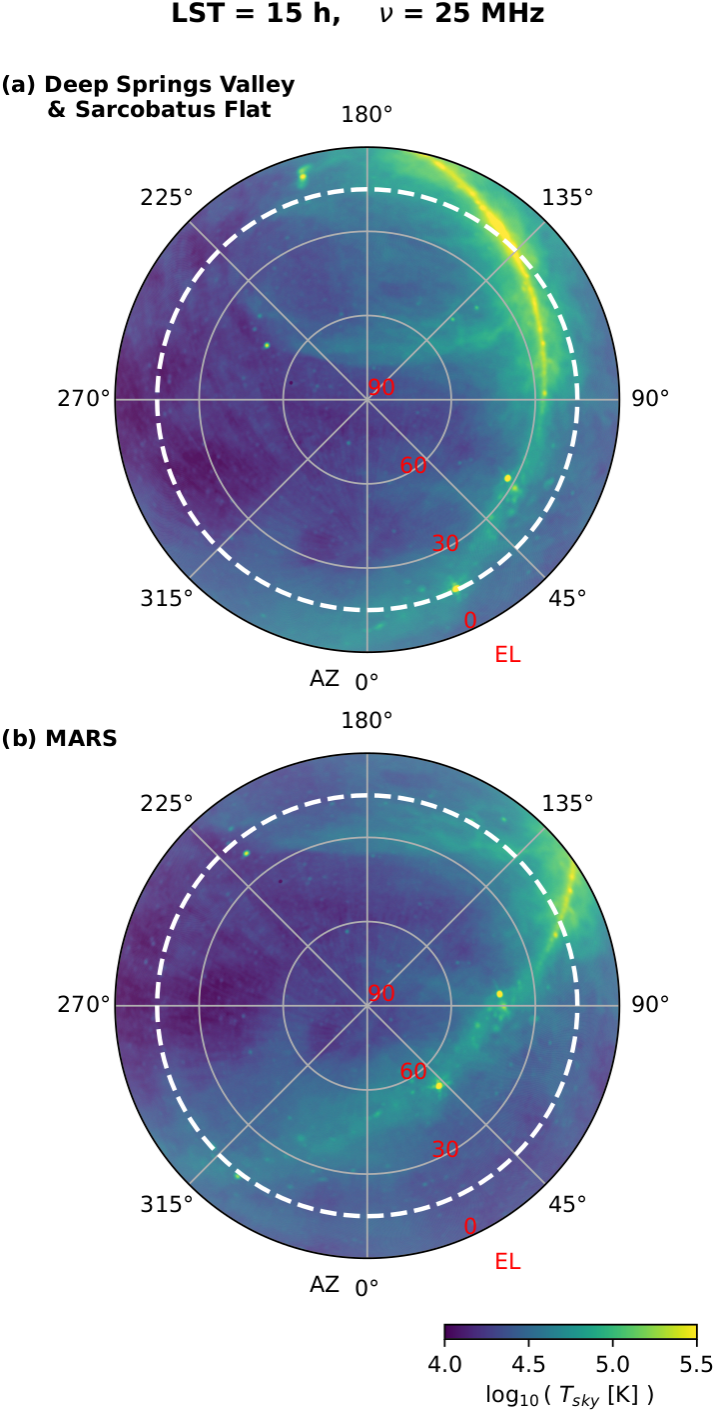}
\caption{View of the sky at LST=$15$~h and $25$~MHz from: (a) Deep Springs Valley and the Sarcobatus Flat, both of which are at a latitude of $\approx37.3^{\circ}$~N; and (b) MARS, at a latitude of $\approx79.38^{\circ}$~N. For this figure, we used the Global Sky Model \citep{de_oliveira_costa2008, price2016}. The dashed lines represent the $-10$~dB contours of the beam directivity at $25$~MHz for the nominal soil model, shown as reference.}
\label{figure_LST15}
\end{figure}

\section{Field measurements}
\label{section_field_measurements}

Between May and July 2022, we conducted sky measurements with MIST in Deep Springs Valley in California ($37.34583^{\circ}$~N, $118.02555^{\circ}$~W), the Sarcobatus Flat in Nevada ($37.21333^{\circ}$~N, $117.09111^{\circ}$~W), and MARS in the Canadian High Arctic ($79.37980^{\circ}$~N, $90.99885^{\circ}$~W). The sites in the Deep Springs Valley and the Sarcobatus Flat were used for observations because they offer a relatively low-RFI environment \citep{monsalve2014,anderson2016} and flat soil while remaining close to urban areas. These sites are, respectively, $\approx$20~km and $\approx$100~km east of the Owens Valley Radio Observatory (OVRO) \citep{daddario2019}. MARS is located $>1800$~km away from cities and possesses an extremely low-RFI environment \citep{dyson2021}.

Figures~\ref{figure_field} and \ref{figure_field2} show short sample measurements from the three sites. These data preliminarily indicate that MIST is performing as expected. A more in-depth discussion of instrument performance, observing sites, and analysis of the full data sets will be presented in future papers.

\subsection{Power spectral densities}
\label{section_field_psd}
Each column of Figure~\ref{figure_field} corresponds to one of the three sites. Instrument~$1$ was used at Deep Springs, and instrument~$2$ was used at the Sarcobatus Flat and MARS. The top row of the figure shows raw, unprocessed PSDs from the antenna, the internal ambient load, and the ambient load plus noise source. Each of the PSDs is a $10$-s integration. In each panel, the internal measurements belong to the same PSD cycle as the antenna measurement, which was taken around 15:00 local sidereal time (LST). The data shown from Deep Springs and the Sarcobatus Flat were taken at night, while those from MARS are from the continuous daytime of the Arctic summer.  We have not excised RFI from these measurements.

\subsection{Reflection coefficient at balun output} 
The second row in Figure~\ref{figure_field} shows the calibrated reflection coefficient of the antenna at the balun output. This quantity was measured at the beginning of the same $111$-min measurement block as the PSDs. We see that the magnitude and phase have the same general shape as those shown in Figure~\ref{figure_impedance_balun}, which were obtained from electromagnetic simulations propagated through the balun S-parameters. One feature that differentiates the three sites is the magnitude dip at $\approx50$~MHz. This dip reaches $\approx-25$, $-16$, and $-20$~dB, at Deep Springs Valley, the Sarcobatus Flat, and MARS, respectively. The reflection coefficient is sensitive to the soil parameters, as discussed in Sections~\ref{section_antenna_impedance} and \ref{section_balun_reflection_coefficient}. Taking advantage of this sensitivity, the soil properties at the different observing sites can be estimated by fitting electromagnetic simulations to the measured reflection. This analysis will be presented in future work.

\subsection{Receiver gain and temperature}
The third row in Figure~\ref{figure_field} shows the receiver gain and temperature for the three observations. The receiver gain and temperature were computed using Equations~\ref{equation_full_receiver_gain} and \ref{equation_full_receiver_temperature}, respectively, and the PSDs from the internal calibrators shown in the first row. The gain (temperature) is relatively stable across the frequency range but quickly decreases (increases) toward lower frequencies below $\approx45$~MHz due to the two $48$~MHz high-pass filters used in the amplification chain to suppress shortwave RFI. The gain and temperature are similar between the two instruments and three observations, with differences $<10\%$.

\subsection{Receiver calibration, and radiation and balun efficiency correction}
\label{section_field_antenna_temperature}
The bottom row of Figure~\ref{figure_field} shows the antenna temperature spectrum obtained from the PSD after applying: (1) receiver calibration, and (2) receiver calibration plus radiation and balun efficiency corrections. Applying the receiver calibration consists of applying to the antenna PSD the receiver gain and temperature shown in the third row of the figure following Equation~\ref{equation_temperature_psd}. The receiver gain and temperature are computed using the measurements of the internal calibrators from the same PSD cycle. The effective sky time for each antenna PSD is $10$~s but because the PSD cycles have a duration of $41$~s (Table~\ref{table_measurement_block}), we refer to the antenna temperature spectra after receiver calibration as $41$-s integrations.

The antenna temperature with only receiver calibration and still affected by all the losses is significantly lower than expected from the astrophysical foreground in the northern hemisphere \citep{bernardi2016,dowell2017, dowell2018, spinelli2021}. Furthermore, the antenna temperature has a peak at $\approx45$--$50$~MHz and a decrease toward lower frequencies, departing from the power law that characterizes the foreground. This departure is mainly caused by the balun efficiency which, as Figure~\ref{figure_balun3} shows, decreases quickly toward lower frequencies. After applying the radiation and balun efficiency corrections the antenna temperature increases significantly, especially at low frequencies, and its shape becomes closer to the expected power law. Since the radiation efficiency is $>99.9\%$ (Figure~\ref{figure_radiation_efficiency}) while the balun efficiency peaks at $\approx45\%$ (Figure~\ref{figure_balun3}), the balun efficiency correction dominates the combined correction.

\subsection{Beam efficiency correction}
\label{section_beam_efficiency_correction}
Figure~\ref{figure_field2} shows an example of the beam efficiency correction. For this example we use the nominal beam efficiency, i.e. the beam efficiency produced by the FEKO simulation that uses the nominal single-layer soil model. As Figure~\ref{figure_beam_efficiency} shows, the nominal beam efficiency ranges between $11\%$ at $25$~MHz and $56\%$ at $105$~MHz. This curve is not expected to represent well the soil properties at our observation sites.  Measuring soil electrical properties to inform more accurate, site-specific beam efficiency corrections is work in progress and will be discussed in future papers.

Figure~\ref{figure_field2} shows that, after beam efficiency correction, the antenna temperature from the three sites increases significantly and its shape becomes closer to the power law expectation. The sites in Deep Springs Valley and the Sarcobatus Flat are at a similar latitude ($\approx37.3^{\circ}$~N). Therefore, at the same LST these sites have a similar visible sky and sky-averaged contribution to the antenna temperature. The difference seen in Figure~\ref{figure_field2} between the corrected spectra from these two sites is mainly due to different beam efficiencies in the observations resulting from different soil characteristics. The spectrum from Deep Springs Valley, in particular, still shows structure that deviates from a power law. This structure is an indication that the soil at this site departs in important ways from the nominal soil model used for the figure. 

The MARS spectrum differs from the others because of the site's contrasting soil characteristics and significantly higher latitude ($\approx 79.38^{\circ}$~N). Figure~\ref{figure_LST15} shows a view of the sky at $25$~MHz and LST=$15$~h from both latitudes. The fraction of the Galactic plane that is visible from Deep Springs Valley and the Sarcobatus Flat is larger than from MARS.

\subsection{Preliminary spectral index}
As a preliminary characterization of the calibrated and efficiency corrected spectra presented in Section~\ref{section_beam_efficiency_correction}, here we compute their spectral index. Since the beam efficiency correction applied as an example in Section~\ref{section_beam_efficiency_correction} corresponds to the nominal soil model, the accuracy of the spectral indices is not expected to be high. Furthermore, we have not removed from the data the effect produced by the frequency dependence of the beam directivity, i.e. the `beam chromaticity'. The beam chromaticity changes the spatial weighting of the sky brightness temperature as a function of frequency and, thus, can have a significant impact on the spectral index \cite[e.g.,][]{mozdzen2017,mozdzen2019, spinelli2021}. As discussed in Section~\ref{section_beam_directivity}, the beam directivity is very sensitive to the properties of the soil, highlighting again that soil characterization a critical task for MIST. We will address this aspect in future analyses.

Instead of fitting functions to the data, we take a simpler approach and compute the spectral index across our frequency range as

\begin{equation}
\beta = \frac{\log\left(\frac{T_S(105\;\mathrm{MHz})}{T_S(25\;\mathrm{MHz})}\right)}{\log\left(\frac{105\;\mathrm{MHz}}{25\;\mathrm{MHz}}\right)},
\end{equation}

\begin{table}
\caption{Spectral indices for the sample spectra from the three observation sites. Beam efficiency correction has been applied but assuming the nominal soil model. Beam chromaticity correction has not been applied. The antenna temperatures shown and used for the computation of $\beta$ are representative values within the measurement noise of the $41$-s integrated spectra at $25$ and $125$~MHz.}             % title of Table
\label{table_beta}      % is used to refer this table in the text
\centering                          % used for centering table
\begin{tabular}{l c c c}        % centered columns (4 columns)
\hline % inserts double horizontal lines
\\
 Site & $T_S(25~\mathrm{MHz})$ & $T_S(105~\mathrm{MHz})$ & $\beta$ \\ % table heading 
\hline  
Deep Springs Valley & 60000~K  & 1100~K  &  $-2.79$ \\
Sarcobatus Flat     & 90000~K  & 1150~K  &  $-3.04$ \\
MARS                & 35000~K  &   950~K  &  $-2.51$ \\
\hline                                   %inserts single line
\end{tabular}
\end{table}

\noindent where $T_S$ is the calibrated and efficiency corrected sky spectrum.

Table~\ref{table_beta} shows the antenna temperatures and spectral index for the three sites. For Deep Springs Valley, the Sarcobatus Flat, and MARS, the spectral indices are $-2.79$, $-3.04$, and $-2.51$, respectively. These values are in broad agreement with previous results for the Northern hemisphere \cite[e.g.,][]{dowell2018,spinelli2021}. This agreement is a good indication for MIST, and future refinements of the beam efficiency, as well as beam chromaticity correction, should improve the results.

\section{Summary and conclusions}
\label{section_summary}

In this paper, we have provided an overview of the two instruments built so far as part of the MIST global $21$~cm ground-based experiment. The instruments are single-antenna, single-polarization, total-power radiometers measuring the sky in the range $25$--$105$~MHz. For the $21$~cm signal, this range corresponds to redshifts $55.5>z>12.5$, which encompasses the Dark Ages and Cosmic Dawn. 

Distinctive characteristics of MIST compared with the existing instruments are: 

\begin{enumerate}[(1)]

\item MIST operates above the soil but without a metal ground plane. This choice was made to avoid systematic effects from the ground plane and its interaction with the antenna. Not using a ground plane results in a higher ground loss and overall dependence of the antenna performance on the soil's electrical characteristics. However, this choice defines an instrumental parameter space that interacts with the $21$~cm signal differently from current experiments.

\item The instruments have been designed with high portability and compactness in mind in order to conduct measurements from remote sites. Before assembly, the largest parts of the instrument are the $3$-mm thick, $1.2$~m~$\times$~$0.6$~m antenna panels. The instruments are battery-powered and have a maximum power consumption of only $17$~W. The batteries and all the electronics apart from the balun are contained in a single metal receiver box located under the antenna. This compactness eliminates systematics related to cables near the antenna and is a unique aspect of MIST, contrasting with other experiments where the back-end or auxiliary electronics are kept at a secondary location. 

\item The MIST observation procedure includes measurements of the antenna reflection coefficient and the reflection coefficient looking into the receiver input. These reflection measurements are done every $111$~min with a VNA integrated into the instrument. The antenna reflection measurements, in addition to being used for receiver calibration, will be used to precisely determine the electrical parameters of the soil. This represents another unique aspect of MIST.

\item Thanks to its portability and low power consumption, MIST has already conducted observations from Deep Springs Valley in California and the Sarcobatus Flat in Nevada, as well as from MARS at a latitude of $\approx 79.38^{\circ}$~N. MARS, in particular, offers excellent radio-quiet conditions and a view of the sky that differs from, and thus complements, observations from lower latitudes by MIST and other experiments.

\end{enumerate}

We quantified the performance of the MIST antenna using electromagnetic simulations and showed results for the beam directivity, radiation efficiency, beam efficiency, and reflection coefficient. Through examples, we have described how these parameters depend on the soil characteristics. As expected, the parameters are more sensitive to soil changes that occur closer to the surface, but changes below the surface are also significant and produce spectral ripples in the parameters. We discuss the impact of the MIST beam directivity on the global $21$~cm signal extraction for different soil characteristics in \citet{monsalve2024}.

We have shown sample sky measurements taken in 2022 with the two MIST instruments from Deep Springs Valley, the Sarcobatus Flat, and MARS. These measurements are short integrations with different levels of calibration, including a preliminary correction for beam efficiency that should improve in the future after an accurate determination of the soil electrical parameters. The measurements preliminarily show that the instruments have the expected performance in the field. We computed initial estimates for the sky spectral index and they were found in the range $\approx -2.5$ to $\approx -3$, consistent with previous measurements \citep{dowell2018,spinelli2021}. These values will be refined after improving the beam efficiency correction and applying beam chromaticity correction. We leave for future work the detailed analysis of the data taken in 2022.

Taking advantage of MIST’s portability, we plan to carry out joint analyses leveraging observations from different latitudes over different soils to help separate the $21$~cm signal from other spectral contributions. This strategy is conceptually similar to some suggested in previous works where, for instance, data from more than one antenna, sidereal time, or Stokes polarization component, are used to improve the signal separation \citep{nhan2019,tauscher2020,deleraacedo2022,anstey2023,saxena2023}.

\section*{acknowledgements}
We are grateful to the anonymous reviewer for contributions that significantly improved this paper. We are grateful to Stuart Bale, Judd Bowman, Nivedita Mahesh, Steven Murray, Alan Rogers, Kaja Rotermund, Benjamin Saliwanchik, Peter Sims, An\v{z}e Slosar, and Aritoki Suzuki for useful discussions on instrumentation for $21$~cm cosmology. We are grateful to Matt Dobbs for lending us equipment from his Cosmology Instrumentation Laboratory at McGill University. We are grateful to Brian Hill, Sue Darlington, and Emily Rivera for welcoming us at the Deep Springs College and providing logistical support during the measurements at Deep Springs Valley. We are grateful to Laura Thomson and Christopher Omelon for welcoming us and providing logistical support at the McGill Arctic Research Station. We are also grateful to Jo\"{e}lle Begin, Cherie Day, Eamon Egan, Larry Herman, Marc-Olivier Lalonde, and Tristan M\'enard for their support with operations in the Arctic. We also acknowledge the Polar Continental Shelf Program for providing funding and logistical support for our research program, and we extend our sincere gratitude to the Resolute staff for their generous assistance and bottomless cookie jars. We acknowledge support from ANID Chile Fondo 2018 QUIMAL/180003, Fondo 2020 ALMA/ASTRO20-0075, Fondo 2021 QUIMAL/ASTRO21-0053. We acknowledge support from Universidad Cat\'olica de la Sant\'isima Concepci\'on Fondo UCSC BIP-106. We acknowledge the support of the Natural Sciences and Engineering Research Council of Canada (NSERC), RGPIN-2019-04506, RGPNS 534549-19. We acknowledge the support of the Canadian Space Agency (CSA) [21FAMCGB15]. This research was undertaken, in part, thanks to funding from the Canada 150 Research Chairs Program. This research was enabled in part by support provided by SciNet and the Digital Research Alliance of Canada.

\section*{Data availability}
The data underlying this article will be shared on reasonable request to the corresponding author.

% Don't change these lines
\bsp	% typesetting comment
\label{lastpage}
\end{document}